\documentclass{emulateapj}

\usepackage{graphicx}
\usepackage{datetime}

\newcommand{\cf}{{\ifmmode{C_{\rm f}}\else{$C_{\rm f}$}\fi}}
\newcommand{\zl}{{\ifmmode{z_{\rm l}}\else{$z_{\rm l}$}\fi}}
\newcommand{\zem}{{\ifmmode{z_{\rm em}}\else{$z_{\rm em}$}\fi}}
\newcommand{\zabs}{{\ifmmode{z_{\rm abs}}\else{$z_{\rm abs}$}\fi}}
\newcommand{\kms}{{\ifmmode{{\rm km~s}^{-1}}\else{km~s$^{-1}$}\fi}}
\newcommand{\cm}{{\ifmmode{{\rm cm}^{-1}}\else{cm$^{-1}$}\fi}}
\newcommand{\cmm}{{\ifmmode{{\rm cm}^{-2}}\else{cm$^{-2}$}\fi}}
\newcommand{\cmmm}{{\ifmmode{{\rm cm}^{-3}}\else{cm$^{-3}$}\fi}}
\newcommand{\lya}{{\rm Ly}$\alpha$} 
\newcommand{\dtra}{{\ifmmode{D_{\rm tra}}\else{$D_{\rm tra}$}\fi}}
\newcommand{\rone}{{\ifmmode{R_{\rm 1on}}\else{$R_{\rm 1on}$}\fi}}
\newcommand{\dew}{{\ifmmode{{\rm d}EW}\else{d$EW$}\fi}}
\newcommand{\rew}{{\ifmmode{REW}\else{$REW$}\fi}}
\newcommand{\rews}{{\ifmmode{REW{\rm s}}\else{$REW{\rm s}$}\fi}}
\newcommand{\ew}{{\ifmmode{EW}\else{$EW$}\fi}}
\newcommand{\ews}{{\ifmmode{EW{\rm s}}\else{$EW{\rm s}$}\fi}}

\newcounter{species} 
\def\ion#1#2{\setcounter{species}{#2}#1$\;${\scriptsize\Roman{species}}\relax}

\usepackage{color}

\slugcomment{Draft version \usdate\today}

\shorttitle{Multi-Sightline Observation of CGM}
\shortauthors{Koyamada et al.}

\begin{document}

\title{Resolving the Internal Structure of Circum-Galactic Medium
  using Gravitationally Lensed Quasars$^{1}$}

\author{Suzuka Koyamada$^{2}$,
        Toru Misawa$^{3}$,
        Naohisa Inada$^{4}$,
        Masamune Oguri$^{5,6,7}$,
        Nobunari Kashikawa$^{8,9}$,
        Katsuya Okoshi$^{10}$}

\affil{$^{1}$Based on data collected at Subaru Telescope, which isoperated by the National Astronomical Observatory of Japan.}
\affil{$^{2}$Department of Physics, Faculty of Science, Shinshu
  University, 3-1-1 Asahi, Matsumoto, Nagano 390-8621 Japan, e-mail:szk.koyamada@gmail.com}
\affil{$^{3}$School of General Education, Shinshu University,
  3-1-1 Asahi, Matsumoto, Nagano 390-8621, Japan}
\affil{$^{4}$Department of Physics, Nara National College of
  Technology, Yamatokohriyama, Nara 639-1080, Japan}
\affil{$^{5}$Research Center for the Early Universe, University of
  Tokyo, 7-3-1 Hongo, Bunkyo-ku, Tokyo 113-0033, Japan}
\affil{$^{6}$Department of Physics, University of Tokyo, 7-3-1
  Hongo, Bunkyo-ku, Tokyo 113-0033, Japan} 
\affil{$^{7}$Kavli Institute for the Physics and Mathematics of
  the Universe (Kavli IPMU, WPI), University of Tokyo, Chiba 277-8583,
  Japan}
\affil{$^{8}$Department of Astronomy, School of Science, SOKENDAI
  (The Graduate University for Advanced Studies), Mitaka, Tokyo
  181-8588, Japan}
\affil{$^{9}$Optical and Infrared Astronomy Division, National
  Astronomical Observatory of Japan, Mitaka, Tokyo 181-8588, Japan}
\affil{$^{10}$Tokyo University of Science, 102-1 Tomino,
  Oshamambe, Hokkaido, 049-3514, Japan}


\begin{abstract}
We study the internal structure of the Circum-Galactic Medium (CGM),
using 29 spectra of 13 gravitationally lensed quasars with image
separation angles of a few arcseconds, which correspond to 100~pc to
10~kpc in physical distances. After separating metal absorption lines
detected in the spectra into high-ions with ionization parameter (IP)
$>$ 40~eV and low-ions with IP $<$ 20~eV, we find that i) the fraction
of absorption lines that are detected in only one of the lensed images
is larger for low-ions ($\sim$16\%) than high-ions ($\sim$2\%), ii)
the fractional difference of equivalent widths (\ews) between the
lensed images is almost same (\dew\ $\sim$ 0.2) for both groups
although the low-ions have a slightly larger variation, and iii) weak
low-ion absorbers tend to have larger \dew\ compared to weak high-ion
absorbers.  We construct simple models to reproduce these observed
properties and investigate the distribution of physical quantities
such as size and location of absorbers, using some free parameters.
Our best models for absorbers with high-ions and low-ions suggest that
i) an overall size of the CGM is at least $\sim$ 500~kpc, ii) a size
of spherical clumpy cloud is $\sim$ 1~kpc or smaller, and iii) only
high-ion absorbers can have diffusely distributed homogeneous
component throughout the CGM. We infer that a high ionization absorber
distributes almost homogeneously with a small-scale internal
fluctuation, while a low ionization absorber consists of a large
number of small-scale clouds in the diffusely distributed higher
ionized region. This is the first result to investigate the internal
small-scale structure of the CGM, based on the large number of
gravitationally lensed quasar spectra.
\end{abstract}

\keywords{galaxies:formation -- intergalactic medium}

\section{INTRODUCTION}
Cosmologically intervening metal absorbers detected in spectra of
background quasars \citep[e.g.,][]{lan90,ber91} and galaxies
\citep[e.g.,][]{ade05,ste10} are good probes of the Circum-Galactic
Medium (CGM) of foreground galaxies. The CGM, which is fuel for star
formation in the galaxy and/or ejected matter blown out by galactic
winds, recently attracts a lot of attention as it is a key ingredient
to understand galaxy formation and evolution.  Based on multiple
galaxy-CGM spectroscopy, the radial gradient of equivalent width (\ew)
and column density ($\log{N}$) of both hydrogen and metal absorbers in
the CGM as a function of transverse and line-of-sight directions were
built up to several proper Mpc (pMpc, hereafter) from galaxies at
\zabs\ $<$ 0.5 \citep[e.g.,][]{che10a,che10b,tum11,pro11,wer16} and at
\zabs\ $\sim$ 2 -- 3 \citep[e.g.,][]{rak12,rak13,tur14,tur15,rub15}.
Several studies have revealed the radial gradient of physical
conditions in the CGM such as covering factor (\cf), ionization
parameter ($\log{U}$), gas temperature ($T_{\rm gas}$), and turbulence
velocity ($v_{\rm turb}$)
\citep[e.g.,][]{rud12,rak12,pro13,tur14,lau15}.  Recently, projected
2D maps along our sightline have also been built through
multi-sightline spectroscopy \citep{pro14,ste16}, deep narrow-band
imaging \citep{can14,hen15} and integral field spectroscopy
\citep[e.g.,][]{bor16}.  \citet{arr16} suggested that diffuse
\lya\ emission from the CGM could distribute up to $\sim$ 500 proper
kpc (pkpc, hereafter) from quasar host galaxies with a detection limit
of SB$_{{\rm Ly}\alpha}$ = $5.5 \times 10^{-20}$
erg~s$^{-1}$~cm$^{-2}$~arcsec$^{-2}$.  Thus, a {\it global} picture of
the CGM is being formed progressively.  On the other hand, very little
has been known about an {\it internal} small scale structure of the
CGM, such as i) their homogeneity or clumpiness, and ii) a typical
scale of each clumpy cloud (or density fluctuation) if the latter is
the case.  One of the difficulty to probe the CGM internal structure
is that only one-dimensional distribution can basically be drawn using
a background spectrum.

\begin{deluxetable*}{ccccccccc}
\tabletypesize{\tiny}
\tablecaption{Sample Quasars\label{t1}}
\tablewidth{0pt}
\tablehead{
\colhead{Lensed QSO}         &
\colhead{\zem$^a$}      &
\colhead{\zl$^b$}      &
\colhead{$\theta$$^c$}       &
\colhead{Instrument}         &
\colhead{$\lambda$-coverage} &
\colhead{$\lambda$/$\Delta\lambda$$^d$}            &
\colhead{$T_{\rm exp}$}        &
\colhead{Reference$^e$}      \\
\colhead{}                    &
\colhead{}                    &
\colhead{}                    &
\colhead{(arcsec)}            &
\colhead{}                    &
\colhead{(\AA)}               & 
\colhead{}                    &
\colhead{(sec)}               &
\colhead{}                    \\
\colhead{(1)}            &
\colhead{(2)}            &
\colhead{(3)}            &
\colhead{(4)}            &
\colhead{(5)}            &
\colhead{(6)}            &
\colhead{(7)}            &
\colhead{(8)}            &
\colhead{(9)}            
}
\startdata
SDSS~J024634.11$-$082536.2 & 1.686 &      0.724 & 1.04             & Keck/ESI     & 3900 -- 11000 & $\sim$ 27000 &  900 & 1 \\
SDSS~J074653.03+440351.3   & 1.998 &      0.513 & 1.08             & Keck/ESI     & 3900 -- 11000 & $\sim$ 27000 & 1200 & 2 \\
SDSS~J080623.70+200631.9   & 1.538 &      0.573 & 1.49             & Keck/ESI     & 3900 -- 11000 & $\sim$ 27000 &  900 & 3 \\
SDSS~J090404.15+151254.5   & 1.826 & $\sim$ 0.30 & 1.13             & Gemini/GMOS  & 3700 -- 9800  & $\sim$ 1000  & 4200 & 4 \\
SDSS~J092455.87+021924.9   & 1.523 &      0.394 & 1.81             & Keck/ESI     & 3900 -- 11000 & $\sim$ 27000 & 1200 & 5 \\
SDSS~J100128.61+502756.8   & 1.841 &      0.415 & 2.86             & Gemini/GMOS  & 3700 -- 9800  & $\sim$ 1000  & 4800 & 6 \\
SDSS~J113157.72+191527.7   & 2.915 & $\sim$ 0.30 & 1.46             & Gemini/GMOS  & 3700 -- 9800  & $\sim$ 1000  & 4800 & 4 \\
SDSS~J125819.24+165717.6   & 2.702 &      0.505 & 1.28             & Gemini/GMOS  & 3700 -- 9800  & $\sim$ 1000  & 4800 & 7 \\
SDSS~J134929.84+122706.8   & 1.722 & $\sim$ 0.65 & 3.00             & Gemini/GMOS  & 3700 -- 9800  & $\sim$ 1000  & 3600 & 4 \\
SDSS~J135306.35+113804.7   & 1.624 & $\sim$ 0.25 & 1.41             & Keck/ESI     & 3900 -- 11000 & $\sim$ 27000 &  600 & 3 \\ \hline
HE1104$-$1805                & 2.319 &       0.73 & 3.19             & 3.9m-AAT/RGO & 3170 -- 7570  & $\sim$ 12000 & $\sim$ 5200  & 8 \\
H1413+1143                 & 2.551 & $\sim$ 1.88 & 0.76, 0.86, 1.10 & HST/FOS      & 3250 -- 6500  & $\sim$ 1300  & $\sim$ 4600  & 9 \\
APM08279+5255              & 3.911 &       1.06 & 0.15, 0.38       & HST/STIS     & 5970 -- 8600  & $\sim$ 5000  & $\sim$ 14900 & 10 
\enddata
\tablenotetext{a}{Quasar emission redshift.}
\tablenotetext{b}{Redshift of a foreground lensing galaxy. Approximate values are photometric redshifts rather than spectroscopic redshifts.}
\tablenotetext{c}{Separation angle between lensed images seen from us
  in arcsec.}
\tablenotetext{d}{Spectral resolution}
\tablenotetext{e}{References --- 1: \citealt{ina05}, 2: \citealt{ina07}, 3: \citealt{ina06}, 4: \citealt{kay10}, 5: \citealt{ina03}, 
6: \citealt{ogu05}, 7: \citealt{ina09}, 8: \citealt{sme95}, 9: \citealt{mon98}, and 10: \citealt{ell04}.}
\end{deluxetable*}

Gravitationally lensed quasars are powerful tools for investigating an
internal structure of the CGM. A typical separation angle of lensed
images is a few to tens of arcseconds, corresponding to 100~pc to
100~kpc\footnote[1]{The largest separation distance corresponds to
  large-separation lensed quasars by a cluster of galaxies with a
  separation angle of $\theta$ $>$ 10 arcsec.}  between two paths at
$z$ $\sim$ 1.0 -- 4.0 based on a standard cosmological model.  This
kind of observations have already been partially performed for
investigating cosmologically intervening absorbers
\citep[e.g.,][]{sme95,mon98}.  For example, based on spectra of the
triply imaged quasar APM08279 $+$ 5255, \citet{ell04} suggested an
important trend that : \ion{H}{1} absorbers and high ionization
systems like \ion{C}{4} absorbers show coherence (i.e. coincidence) on
the multiple sightlines over distances of $\sim$100 -- 300~kpc, while
low ionization systems like \ion{Mg}{2} exhibit significant
sightline-variation on scales greater than a few hundred parsec. This
is qualitatively consistent with a simple picture of clumpy, low
ionization gas, embedded in homogeneous, highly ionized outer
halos. These trends are also reminiscent of the hierarchical structure
formation \citep[e.g.,][]{ste16}. However, sample sizes of the past
studies (i.e., only a few lensed-quasars) are not large enough for
statistical analysis.

In this paper, we collected a (large) sample of spectra of
gravitationally lensed quasars to statistically study an internal
structure of the CGM through comparisons of parameters, including
absorption detection rate and rest-frame equivalent widths (\rew) as a
function of ionization condition and physical separation between
lensed images. Because our goal is to resolve the internal small-scale
structure of the CGM, we do not necessarily need to know the positions
of the galaxies hosting the CGM giving rise to the absorption lines
with respect to the background gravitationally lensed quasars.  In
\S2, we describe the data sample and the methods used for detecting
absorption lines and measuring their parameters. The results and
discussion are presented in \S3 and \S4, respectively. We summarize
our results in \S5.  We use a cosmology with $H_{0}$ = 70
\kms~Mpc$^{-1}$, $\Omega_{m}$ = 0.3, and $\Omega_{\Lambda}$ = 0.7
throughout the paper.

\section{DATA and ANALYSIS}

We select our sample quasars from the Sloan Digital Sky Survey Quasar
Lens Search (SQLS; \citealt{ina12}; \citealt{ogu12} and reference
therein).  The SQLS repeatedly performed spectroscopic observations
for gravitationally lensed quasar candidates with various telescopes
and instruments, and discovered 62 lensed quasars. Because the quality
of observed spectra (e.g., wavelength coverage, resolution, and
signal-to-noise (S/N)-ratio) is heterogeneous, we select lensed
quasars whose spectra satisfy all the following criteria: a)
\ion{C}{4} and \ion{Mg}{2} absorption lines are covered by optical
spectra (i.e., quasar emission redshift is larger than 1.5), b)
spectral resolution is grater than 1000, c) wavelength coverage is
wide enough to cover from $\sim$ 4,000 \AA\ to $\sim$ 1 $\mu$m, d)
data quality is high enough (i.e., an S/N-ratio is greater than
$\sim$20 pixel$^{-1}$ on average after sampling in a spectrum).  We
use 20 spectra of 10 lensed quasars taken with Keck/ESI (wavelength
resolution is $\lambda / \Delta \lambda$ $\sim$ 27000) or Gemini/GMOS
($\lambda / \Delta \lambda$ $\sim$ 1000) from SQLS that satisfy the
criteria described above (see Table~\ref{t1}).  Although the spectral
resolution is very different between those taken with Keck/ESI and
Gemini/GMOS, their pixel scale after sampling is almost same,
$\sim$ 1.8~\AA~pix$^{-1}$. We also confirmed the effect of
self-blending (i.e., blue and red members of doublet are blended each
other because of low spectral resolution) is not important when we
measure equivalent width later.\footnote[2]{We synthesized spectra of
  \ion{C}{4} doublet using typical line parameters of \zabs\ = 2.0,
  column density $\log N$ = 14.0~\cmm, and Doppler parameter $b$ =
  50~\kms\ with pixel scale of 1.8~\AA~pix$^{-1}$, S/N ratio of
  20~{pix$^{-1}$}, and two spectral resolutions of $\lambda / \Delta \lambda$ = 
  1000 and 27000. Although the doublet is self-blended only in the 
  $\lambda / \Delta \lambda$ = 1000
  spectrum, we confirmed their total equivalent widths are almost
  same.}  We define the brighter quasar image as image 1, the fainter
as image 2 in optical bands.  Parameters of absorption lines detected
in each lensed image are shown with subscript 1 or 2, hereafter.

In our spectra, we detect all absorption features (except for heavily
blended ones) whose absorption depths at the center are greater than 5
times of noise level using the code {\sc search}
\citep{chu97,chu03}. Then, we identify doublet lines such as
\ion{C}{4}, \ion{Si}{4}, and \ion{Mg}{2} in the spectral region
between \lya\ and the corresponding emission lines. We also search for
other single metal lines at the same redshift as the doublet lines
above.

For all detected doublets, we measure absorption redshifts ($z_{\rm
  1}$, $z_{\rm 2}$) and rest-frame equivalent widths (\rew$_{\rm 1}$,
\rew$_{\rm 2}$) with their 1$\sigma$ uncertainties
($\sigma$(\rew$_{\rm 1}$), $\sigma$(\rew$_{\rm 2}$))
\footnote[3]{This is defined by $\sigma(\rew)$ = $\sqrt{\sum_{i=1}^{N}
    \left(\sigma_i \Delta\lambda \right)^2}$, where $N$ is a number of
  pixels in the absorption profile, $\sigma_i$ is the error in the
  normalized flux at pixel $i$, and $\Delta\lambda$ is the width of
  each pixel in angstrom.}, in the spectra of both lensed images.
Because blue and red members of doublet are sometimes blended each
other especially for the low resolution spectra taken with Gemini/GMOS
($\lambda / \Delta \lambda$ $\sim$ 1000), we calculate {\it total}
\rews\ of two transitions, \ion{C}{4}~$\lambda\lambda$ 1548, 1551
(\ion{C}{4}~1550, hereafter) and \ion{Mg}{2}~$\lambda\lambda$2796,
2803 (\ion{Mg}{2}~2800, hereafter) including both doublet members. We
chose \ion{C}{4} and \ion{Mg}{2} doublets as representative
transitions for high and low-ionization transitions because these
doublets are most frequently detected among each category.  We also
measure above parameters for the other metal lines.

\begin{deluxetable*}{ccccccccccc}
\tabletypesize{\scriptsize}
\tablecaption{Detected Absorption Lines\label{t3}}
\tablewidth{0pt}
\tablehead{
\colhead{Lensed QSO}          &
\colhead{ion}                 &
\colhead{$z_{\rm 1}$}           &
\colhead{\rew$_{\rm 1}$}        &
\colhead{$\sigma$(\rew$_1$)}  &
\colhead{$z_{\rm 2}$}           &
\colhead{\rew$_{\rm 2}$}        &
\colhead{$\sigma$(\rew$_2$)}  &
\colhead{$D_{\rm tra}$}         &
\colhead{\dew}                &
\colhead{$\sigma(\dew)$}      \\
\colhead{}                    &
\colhead{}                    &
\colhead{}                    &
\colhead{(\AA)}                    &
\colhead{(\AA)}               &
\colhead{}               & 
\colhead{(\AA)}               &
\colhead{(\AA)}               & 
\colhead{(pkpc)}  &
\colhead{}                    &
\colhead{}              \\
\colhead{(1)}            &
\colhead{(2)}            &
\colhead{(3)}            &
\colhead{(4)}            &
\colhead{(5)}            &
\colhead{(6)}            &
\colhead{(7)}            &
\colhead{(8)}            &
\colhead{(9)}            &
\colhead{(10)}           &
\colhead{(11)}            
}
\startdata
SDSS~J024634.11$-$082536.2 & \ion{Mg}{2} 2800 &  0.7246 & 0.692 & 0.041 & 0.7256 & 0.978 & 0.068 & 7.86 & 0.292 & 0.064 \\
	& \ion{Mg}{1} 2853 & 0.7242 & 0.284 & 0.037 & 0.7242 & 0.774 & 0.062 & 7.87 & 0.632 & 0.056 \\ \cline{2-11}
	& \ion{Mg}{2} 2800 & 1.1218 & 0.357 & 0.025 & 1.1221 & 0.278 & 0.050 & 3.33 & 0.221 & 0.151 \\ \cline{2-11}
	& \ion{Mg}{2} 2800 & 1.1568 & 0.837 & 0.029 & 1.1572 & 0.897 & 0.051 & 3.04 & 0.066 & 0.062 \\ \cline{2-11}
	& \ion{Mg}{2} 2800 & 1.3534 & 0.818 & 0.025 & 1.3534 & 0.774 & 0.044 & 1.66 & 0.053 & 0.061 \\ \cline{2-11}
	& \ion{C}{4} 1550$^a$ & 1.6892 & 0.128 & 0.018 & --- & $<$0.077 & --- & --- & $>$0.399 & --- \\ \cline{2-11}
	& \ion{C}{4} 1550$^a$ & 1.7323 & 0.432 & 0.024 & 1.7325 & 0.438 & 0.037 & --- & 0.013 & 0.101 \\ \hline
SDSS~J074653.03+440351.3 & \ion{Mg}{2} 2800 & 1.6505 & 0.442 & 0.036 & --- & $<$0.116 & --- & 0.642 & $>$0.738 & --- \\ \cline{2-11}
	& \ion{C}{4} 1550 & 1.9342 & 0.700 & 0.053 & 1.9342 & 0.920 & 0.046 & 0.099 & 0.239 & 0.069 \\ \hline
SDSS~J080623.70+200631.9 & \ion{Fe}{2} 2600 & 0.5736 & 1.530 & 0.123 & 0.5743 & 2.291 & 0.271 & 9.729 & 0.332 & 0.095 \\
	& \ion{Mg}{2} 2800 & 0.5735 & 6.398 & 0.163 & 0.5741 & 5.783 & 0.186 & 9.731 & 0.096 & 0.037 \\
	& \ion{Ca}{2} 3935 & 0.5735 & 0.787 & 0.059 & 0.5739 & 1.021 & 0.072 & 9.733 & 0.229 & 0.079 \\ \hline
SDSS~J090404.15+151254.5 & \ion{Mg}{2} 2800 & --- & $<$0.158 & --- & 0.5516 & 2.139 & 0.114 & 3.242 & $>$0.926 & --- \\ \cline{2-11}
	& \ion{Fe}{2} 2600 & 1.2169 & 0.964 & 0.035 & 1.2170 & 0.557 & 0.065 & 0.886 & 0.422 & 0.071 \\
	& \ion{Mg}{2} 2800 & 1.2170 & 3.850 & 0.035 & 1.2169 & 2.452 & 0.070 & 0.886 & 0.363 & 0.019 \\
	& \ion{Mg}{1} 2853 & 1.2166 & 0.221 & 0.024 & --- & $<$0.157 & --- & 0.887 & $>$0.287 & --- \\ \cline{2-11}
	& \ion{C}{4} 1550 & 1.6130 & 1.041 & 0.049 & 1.6127 & 0.769 & 0.117 & 0.237 & 0.261 & 0.118 \\
	& \ion{Mg}{2} 2800 & 1.6125 & 0.434 & 0.025 & 1.6126 & 0.451 & 0.057 & 0.237 & 0.0374 & 0.134 \\ \cline{2-11}
	& \ion{C}{4} 1550 & 1.6547 & 1.175 & 0.045 & 1.6540 & 1.039 & 0.104 & 0.186 & 0.116 & 0.094 \\
	& \ion{Al}{2} 1670 & 1.6530 & 0.142 & 0.023 & 1.6499 & 0.218 & 0.057 & 0.189 & 0.346 & 0.202 \\
	& \ion{Mg}{2} 2800 & 1.6523 & 0.605 & 0.024 & 1.6526 & 0.268 & 0.048 & 0.188 & 0.557 & 0.081 \\ \cline{2-11}
	& \ion{C}{4} 1550 & 1.7686 & 0.457 & 0.032 & 1.7707 & 0.323 & 0.078 & 0.057 & 0.294 & 0.179 \\ \hline
SDSS~J092455.87+021924.9 & \ion{Mg}{2} 2800 & --- & $<$0.071 & --- & 1.0785 & 0.245 & 0.042 & 1.341 & $>$0.709 & --- \\ \hline
SDSS~J100128.61+502756.8 & \ion{Mg}{2} 2800 & --- & $<$0.358 & --- & 0.4145 & 1.893 & 0.206 & 15.700 & $>$0.811 & --- \\ \cline{2-11}
	& \ion{Fe}{2} 2600 & 0.8723 & 0.606 & 0.021 & 0.8720 & 0.478 & 0.036 & 7.794 & 0.211 & 0.066 \\
	& \ion{Mg}{2} 2800 & 0.8718 & 2.045 & 0.068 & 0.8717 & 1.758 & 0.091 & 7.799 & 0.140 & 0.053 \\
	& \ion{Mg}{1} 2853 & 0.8720 & 0.235 & 0.045 & 0.8719 & 0.219 & 0.058 & 7.797 & 0.068 & 0.303 \\ \cline{2-11}
	& \ion{Si}{2} 1526 & 1.6066 & 0.172 & 0.044 & 1.6065 & 0.275 & 0.058 & 1.103 & 0.372 & 0.208 \\
	& \ion{C}{4} 1550 & 1.6074 & 2.400 & 0.062 & 1.6071 & 2.194 & 0.113 & 1.100 & 0.086 & 0.053 \\
	& \ion{Al}{3} 1854 & 1.6077 & 0.178 & 0.018 & 1.6075 & 0.218 & 0.027 & 1.098 & 0.181 & 0.131 \\
	& \ion{Fe}{2} 2600 & 1.6072 & 0.352 & 0.019 & 1.6071 & 0.379 & 0.019 & 1.100 & 0.072 & 0.068 \\
	& \ion{Mg}{2} 2800 & 1.6074 & 1.559 & 0.023 & 1.6073 & 1.363 & 0.028 & 1.099 & 0.126 & 0.022 \\
	& \ion{Mg}{1} 2853 & 1.6072 & 0.193 & 0.020 & 1.6071 & 0.225 & 0.015 & 1.100 & 0.142 & 0.103 \\ \cline{2-11}
	& \ion{C}{4} 1550 & 1.7542 & 0.644 & 0.041 & 1.7563 & 0.421 & 0.064 & 0.368 & 0.347 & 0.108 \\ \cline{2-11}
	& \ion{Si}{4} 1393$^b$ & 1.7711 & 0.856 & 0.083 & 1.7704 & 1.029 & 0.152 & 0.299 & 0.168 & 0.148\\
	& \ion{C}{4} 1550$^b$& 1.7748 & 5.201 & 0.064 & 1.7742 & 1.738 & 0.078 & 0.282 & 0.666 & 0.016 \\ \cline{2-11}
	& \ion{C}{4} 1550$^c$ & 1.8151 & 1.508 & 0.035 & 1.8147 & 1.076 & 0.046 & 0.108 & 0.286 & 0.035 \\ \hline
SDSS~J113157.72+191527.7 & \ion{Fe}{2} 2600 & 1.1902 & 0.944 & 0.034 & 1.1902 & 1.719 & 0.098 & 1.923 & 0.451 & 0.037 \\
	& \ion{Mg}{2} 2800 & 1.1902 & 2.841 & 0.028 & 1.1902 & 4.577 & 0.080 & 1.924 & 0.379 & 0.012 \\
	& \ion{Mg}{1} 2853 & 1.1902 & 0.082 & 0.017 & 1.1899 & 0.522 & 0.059 & 1.924 & 0.842 & 0.038 \\ \cline{2-11}
	& \ion{Fe}{2} 2600 & 1.4215 & 1.958 & 0.026 & 1.4191 & 2.597 & 0.092 & 1.416 & 0.246 & 0.029 \\ 
	& \ion{Mg}{2} 2800 & 1.4202 & 6.484 & 0.030 & 1.4190 & 5.753 & 0.105 & 1.418 & 0.113 & 0.017 \\
	& \ion{Mg}{1} 2853 & 1.4215 & 0.472 & 0.024 & 1.4189 & 0.707 & 0.081 & 1.416 & 0.333 & 0.084 \\ \cline{2-11}
	& \ion{Mg}{2} 2800 & 1.5615 & 0.578 & 0.027 & --- & $<$0.246 & --- & 1.168 & $>$0.574 & --- \\ \cline{2-11}
	& \ion{Mg}{2} 2800 & 1.7943 & 0.965 & 0.030 & --- & $<$0.301 & --- & 0.837 & $>$0.688 & --- \\ \hline
SDSS~J125819.24+165717.6 & \ion{C}{4} 1550 & 1.8474 & 1.361 & 0.058 & 1.8475 & 1.298 & 0.066 & 1.103 & 0.047 & 0.063 \\ \cline{2-11}
	& \ion{Si}{4} 1393 & 1.9957 & 1.096 & 0.086 & 1.9960 & 1.112 & 0.090 & 0.837 & 0.015 & 0.111 \\
	& \ion{C}{4} 1550 & 1.9973 & 1.159 & 0.053 & 1.9989 & 1.374 & 0.064 & 0.833 & 0.156 & 0.055 \\ \cline{2-11}
	& \ion{C}{4} 1550 & 2.1062 & 1.019 & 0.047 & 2.1066 & 0.855 & 0.056 & 0.664 & 0.161 & 0.067 \\ \cline{2-11}
	& \ion{Si}{4} 1393 & 2.2501 & 0.198 & 0.028 & 2.2505 & 0.147 & 0.024 & 0.466 & 0.259 & 0.158 \\
	& \ion{C}{4} 1550 & 2.2500 & 0.555 & 0.092 & 2.2500 & 0.358 & 0.110 & 0.466 & 0.355 & 0.224 \\ \cline{2-11}
	& \ion{Si}{4} 1393 & 2.3868 & 0.631 & 0.045 & 2.3865 & 0.813 & 0.064 & 0.304 & 0.223 & 0.082 \\
	& \ion{C}{4} 1550 & 2.3848 & 0.521 & 0.066 & 2.3852 & 0.783 & 0.078 & 0.305 & 0.335 & 0.107 \\
	& \ion{Al}{2} 1670 & 2.3842 & 0.344 & 0.036 & 2.3846 & 0.407 & 0.036 & 0.306 & 0.154 & 0.116 \\
	& \ion{Fe}{2} 2600 & 2.3840 & 0.428 & 0.036 & 2.3843 & 0.472 & 0.037 & 0.306 & 0.093 & 0.104 \\ \hline
SDSS~J134929.84+122706.8 & \ion{Mg}{2} 2800 & 0.4913 & 0.936 & 0.083 & --- & $<$0.622 & --- & 18.141 & $>$0.336 & --- \\ \cline{2-11}
	& \ion{Mn}{2} 2576 & --- & $<$0.050 & ---  & 1.2395 & 0.213 & 0.035 & 5.780 & $>$0.765 & --- \\
	& \ion{Fe}{2} 2600 & 1.2374 & 1.559 & 0.032 & 1.2376 & 1.974 & 0.070 & 5.812 & 0.210 & 0.032 \\
	& \ion{Mg}{2} 2800 & 1.2373 & 3.383 & 0.035 & 1.2375 & 4.456 & 0.071 & 5.814 & 0.241 & 0.014 \\
	& \ion{Mg}{1} 2853 & 1.2373 & 0.546 & 0.027 & 1.2375 & 0.817 & 0.058 & 5.814 & 0.332 & 0.058 \\ \hline
SDSS~J135306.35+113804.7 & \ion{Mg}{2} 2800 & 0.6377 & 1.032 & 0.052 & 0.6378 & 0.982 & 0.025 & 2.663 & 0.049 & 0.058 \\ \cline{2-11}
	& \ion{Fe}{2} 2600 & 0.9047 & 0.773 & 0.022 & 0.9048 & 1.263 & 0.009 & 1.537 & 0.387 & 0.018 \\
	& \ion{Mg}{2} 2800$^d$ & 0.9047 & 2.300 & 0.033 & 0.9048 & 3.460 & 0.017 & 1.537 & 0.335 & 0.010 \\
	& \ion{Mg}{1} 2853 & 0.9045 & 0.317 & 0.015 & 0.9048 & 0.292 & 0.045 & 1.538 & 0.080 & 0.149 \\ \cline{2-11}
	& \ion{Fe}{2} 2600 & 1.2386 & 0.204 & 0.022 & 1.2389 & 0.103 & 0.012 & 0.636 & 0.493 & 0.079 \\
	& \ion{Mg}{2} 2800 & 1.2387 & 0.983 & 0.026 & 1.2385 & 0.941 & 0.017 & 0.636 & 0.042 & 0.031 \\ \cline{2-11}
	& \ion{C}{4} 1550 & --- & $<$0.534 & --- & 1.5689 & 0.978 & 0.062 & 0.073 & $>$0.454 & --- 
\enddata
\tablenotetext{a}{Absorption redshift is larger than the quasar emission redshift.}
\tablenotetext{b}{Broad Absorption Line (BAL) with a FWHM $\geq$ 2000~\kms.}
\tablenotetext{c}{Velocity shift from the quasar emission redshift is smaller than 5000~\kms.}
\tablenotetext{d}{This line is blended with other physically unrelated lines.}
\end{deluxetable*}

We also calculate the physical separation in the transverse direction
(\dtra)\footnote[4]{The separation distance between lensed images in
  the transverse direction is calculated by \dtra\ = $\theta D_{\rm
    oa}$ if \zabs\ $<$ \zl\ and \dtra\ = $\theta \frac{D_{\rm
      ol}D_{\rm aq}}{D_{\rm lq}}\frac{1+z_{\rm l}}{1+z_{\rm a}}$ if
  \zabs\ $>$ \zl, where $\theta$ is an angular separation of the
  lensed images seen from us, and the subscripts $_{\rm o}$, $_{\rm
    a}$, $_{\rm l}$, $_{\rm q}$ for $D$ (an angular diameter distance)
  and $z$ denote observer, absorber, lensing galaxy, and quasar,
  respectively. We use an average value of $z_{\rm 1}$ and $z_{\rm 2}$
  for $z_{\rm a}$.}  and the fractional equivalent width difference
\citep{ell04} defined by

\begin{equation}
\dew = \frac{|\rew_{\rm 1} - \rew_{\rm 2}|}{{\rm max}(\rew_{\rm 1}, \rew_{\rm 2})}.
\label{eqn1}
\end{equation}

Because absorption strength is enhanced (compared to the intergalactic
medium) around galaxies up to $\Delta v$ $\sim$ 240~\kms\ along the
line-of-sight for \ion{H}{1} and metal absorption lines including
\ion{C}{4} \citep{tur14}, we assume absorption lines within
400~\kms\ (i.e., $\leq$ 240 $\times$ 2~\kms) each other into a single
absorption ``system''.  As a result, we detected 30 \ion{C}{4}, 8
\ion{Si}{4}, 39 \ion{Mg}{2} doublets as well as 46 single metal lines
in 36 absorption systems in total (see Table~\ref{t3}).

In addition to our data, we also include similar measurements from the
literature for our statistical analysis: double images of HE1104$-$1805
(a quasar emission redshift is \zem\ = 2.319, a redshift of lensing
galaxy is \zl\ = 0.73, and a separation angle is $\theta$ =
3.$^{\!\!\prime\prime}$19) taken with 3.9m-AAT/RGO ($\lambda /
\Delta\lambda$ $\sim$ 12000, $\lambda$ = 3200 -- 7500 \AA;
\citealt{sme95}), quartet images of H1413+1143 (\zem\ = 2.551, \zl\ =
1.88, $\theta$ = 1.$^{\!\!\prime\prime}$10) taken with HST/FOS ($\lambda /
\Delta\lambda$ $\sim$ 1300, $\lambda$ = 3200 -- 6500 \AA; \citealt{mon98}), 
and triple images of APM08279+5255 (\zem\ = 3.911, \zl\ = 1.062, $\theta$ =
0.$^{\!\!\prime\prime}$38) taken with HST/STIS ($\lambda / \Delta\lambda$ 
$\sim$ 5000, $\lambda$ = 6000 -- 8600 \AA; \citealt{ell04}), as shown in
Table~\ref{t1}.

To avoid any possible biases for statistical analysis, we accept only
absorption lines that satisfy all the following criteria: a) they are
blueshifted more than 5000~\kms\ from quasar emission redshifts to
avoid a contamination by absorption lines that are physically
associated to the background quasars, b) they have line widths smaller
than the criterion for broad absorption lines (i.e., 2000~\kms)
because of the same reason as above, c) they are not heavily blended
with other unrelated absorption lines, d) their equivalent width is
larger than 3 times of the noise level (i.e., \rew\ $\geq$
3$\sigma$(\rew)) in spectra\footnote[5]{This is a significance level
  in absorption strength (i.e., equivalent width), while we detected
  above all absorption lines (regardless of absorption strength) based
  on their absorption depth ($>5\sigma$) at the line center.}, and e)
their equivalent width is smaller than 2 \AA\ to avoid a Damped
Ly$\alpha$ (DLA) system whose origin should be different from the CGM.
In Table~\ref{t3}, absorption lines with footnote correspond to
rejected ones. After the above selection, our sample contains 268
metal absorption lines, of which 96 are from our spectra (68 from
Gemini/GMOS and 28 from Keck/ESI spectra), 71 from \citet{ell04}, 60
from \citet{mon98}, and 41 from \citet{sme95}.

Although our study using gravitationally lensed quasars is a
unique and powerful technique to investigate the internal structure
of the CGM, there are several caveats.  For example, in the CGM, a
radial gradient of physical parameter from the center of the
gravitational potential of the galaxy hosting the absorption systems
cannot be investigated because the host galaxies are not identified.
Our sample is also somewhat heterogeneous in terms of spectral
resolution and data quality.  We will discuss these later in \S4.2.

\section{RESULTS}

We classify all absorption lines into two groups; two-on (2on) and
one-on (1on) samples based on line detection with $\geq$
3$\sigma(\rew)$ level in both or one of two spectra of lensed images
within 400~\kms\ from each other.  We also divide absorption lines
into three classes based on their ionization potential (IP): high-ions
with IP $>$ 40~eV (e.g., \ion{N}{5}, \ion{C}{4}, and \ion{Si}{4}),
low-ions with IP $<$ 20~eV (e.g., \ion{Al}{2}, \ion{Ni}{2},
\ion{Si}{2}, \ion{Fe}{2}, \ion{Mn}{2}, \ion{Mg}{2}, \ion{N}{1}, \ion{O}{1},
\ion{Ca}{2}, and \ion{Mg}{1}), and intermediate ones (20~eV $\leq$ IP
$\leq$ 40~eV; e.g., \ion{Si}{3}, \ion{Al}{3}, and \ion{C}{2}). As a
result, we separate 63 high-ions into 59 2on and 4 1on samples, and 99
low-ions into 72 2on and 27 1on samples, respectively.  The rest of
them are absorption lines of intermediately ionized ions.  All
high/low-ionized absorption lines within 400~\kms\ are grouped into
the ``single" absorption system, but most of them have velocity
distributions smaller than 100~\kms.

\begin{figure*}
 \centering
 \includegraphics[width=10cm]{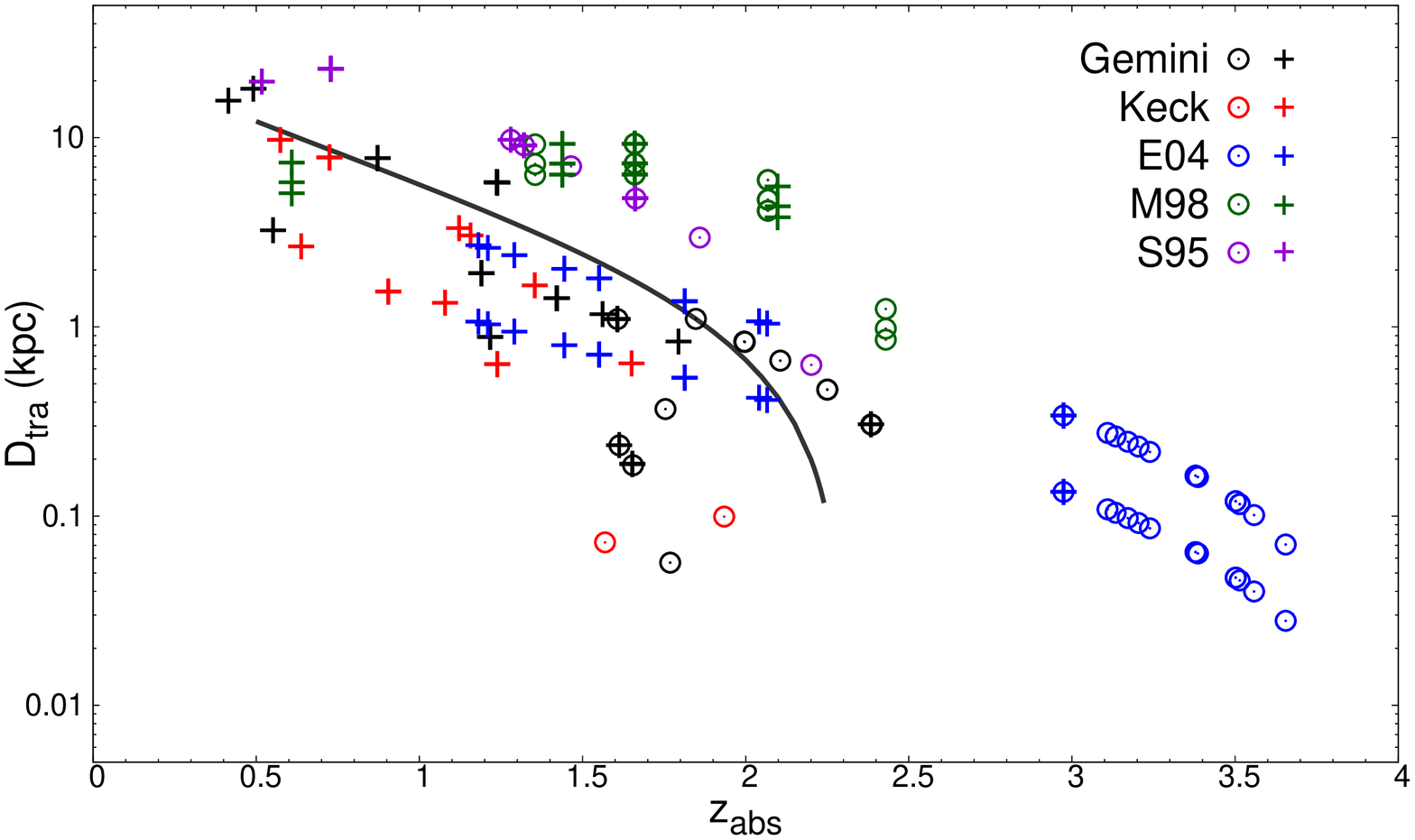}
 \caption{Physical distance between sightlines of lensed images
   (\dtra) as a function of absorption redshift (\zabs). Data from our
   sample are shown with black (Gemini/GMOS) and red (Keck/ESI), while
   those from the literature are shown with blue \citep{ell04}, green
   \citep{mon98}, and purple \citep{sme95}.  open circles and crosses 
   denote high-ion and low-ion absorption lines,
   respectively.  Strong absorption lines with \rew\ $>$ 2 \AA\ are
   also included in the figure.  Thick black curve denotes a physical
   distance between the sightlines corresponding to the typical lensed
   quasar from SQLS (\zem\ = 2.3, \zl\ = 0.5, and $\theta$ =
   2.$^{\!\!\prime\prime}$0). \label{f1}}
\end{figure*}

\begin{figure*}
 \centering
 \includegraphics[width=6cm]{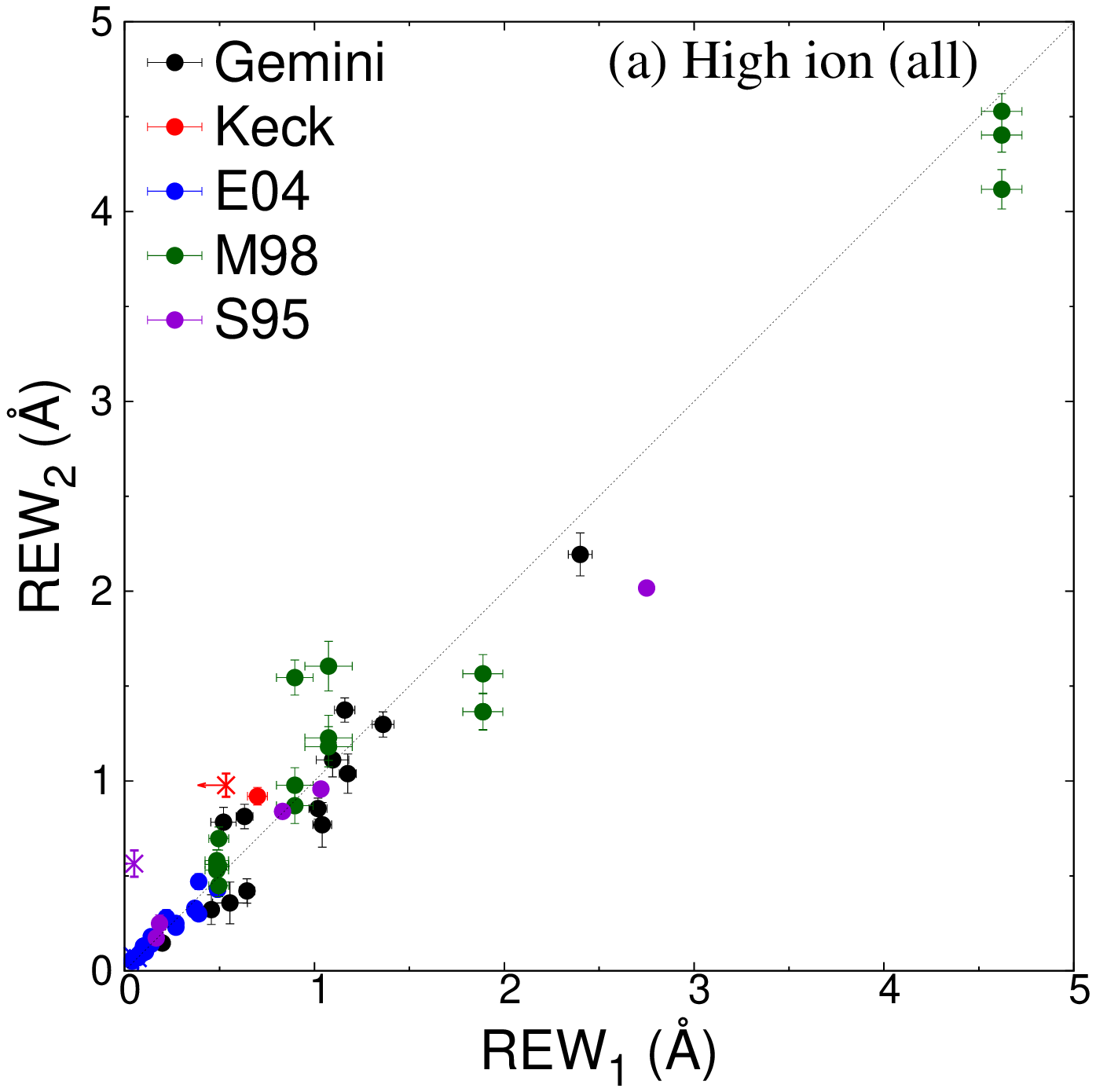}
 \includegraphics[width=6cm]{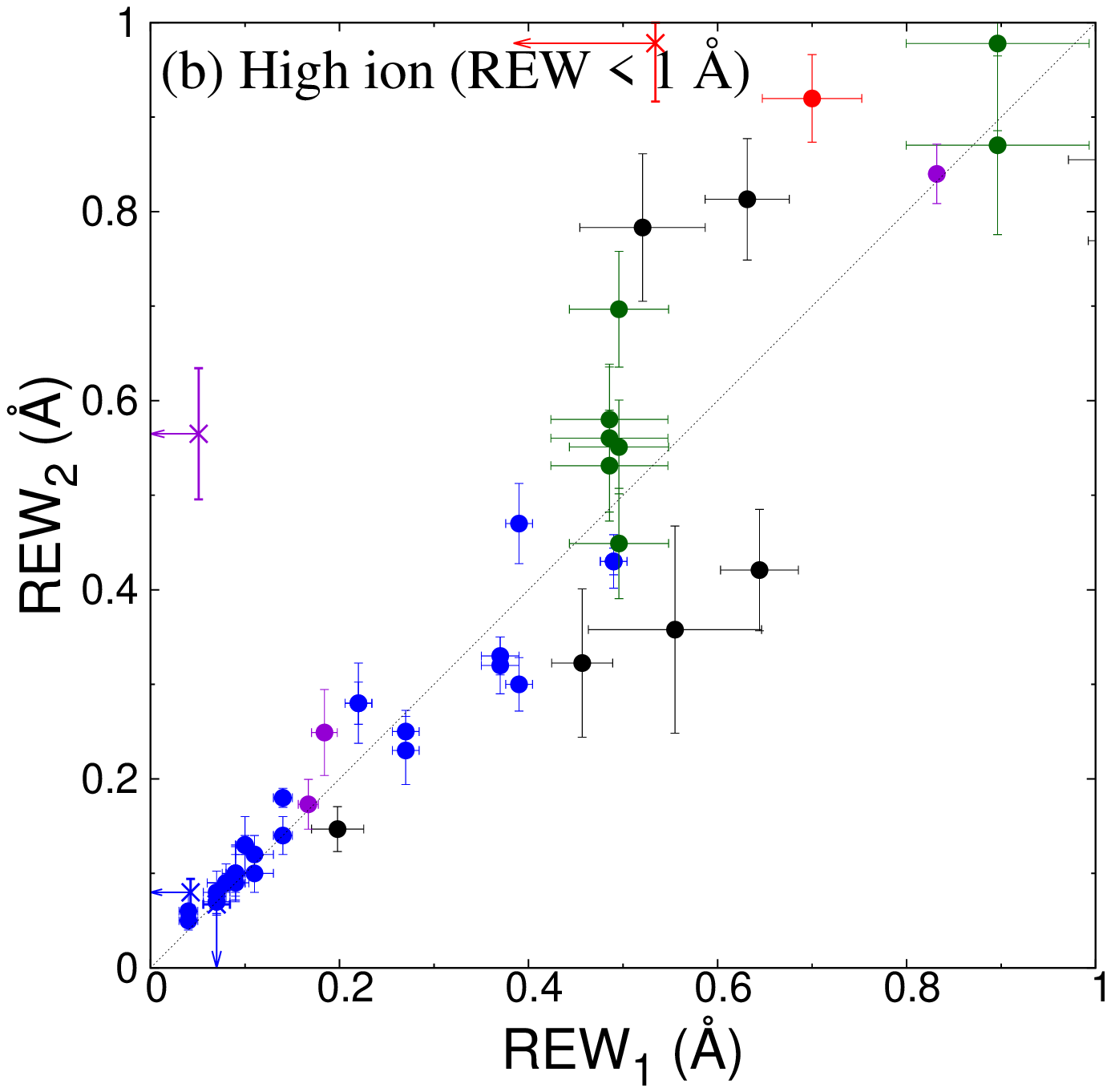}
 \includegraphics[width=6cm]{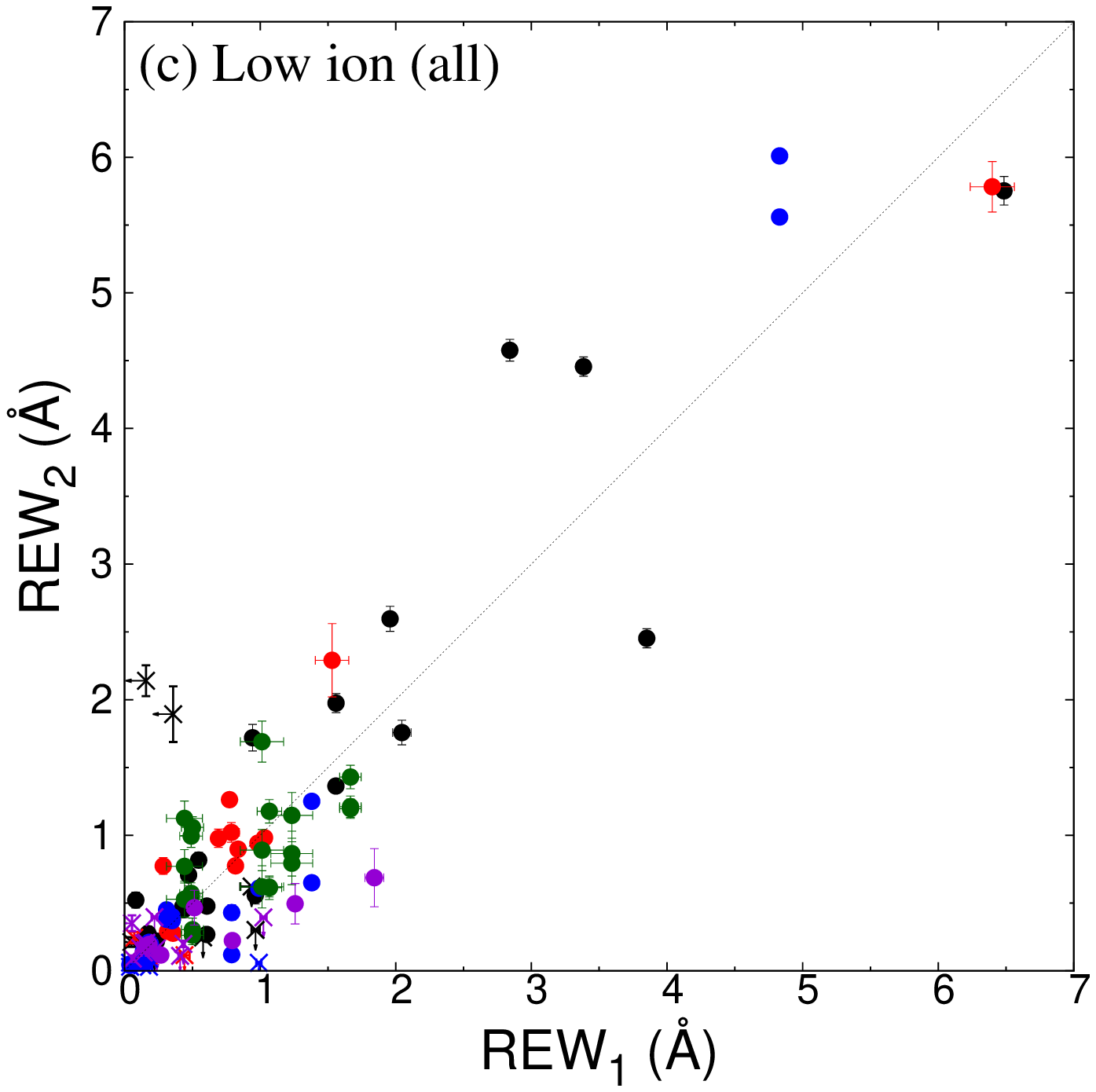}
 \includegraphics[width=6cm]{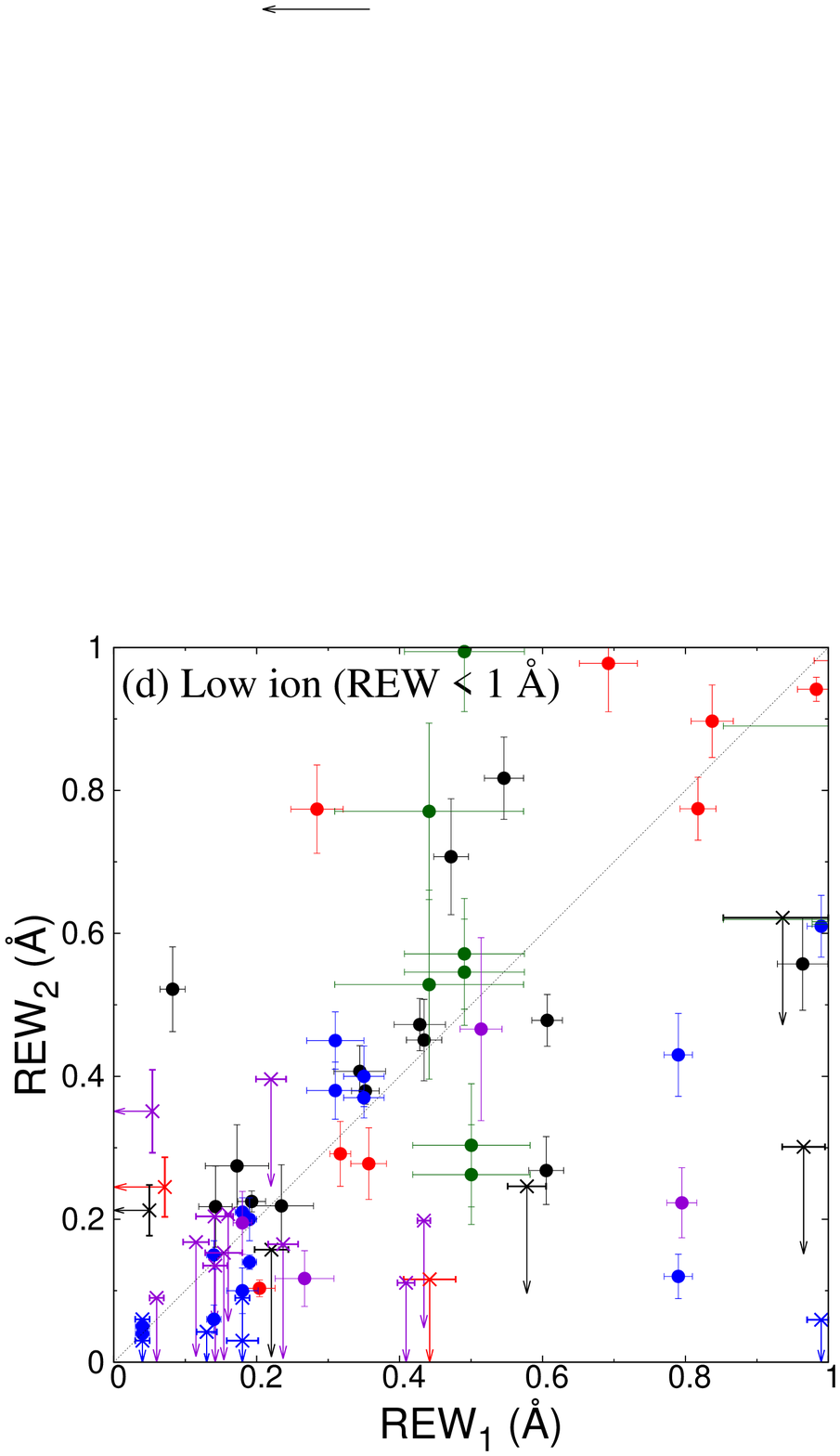}
 \caption{Correlation between rest-frame equivalent widths (\rews) of
   absorption lines detected along both or one of the two sightlines.
   Upper panels show distributions of all high-ion absorption lines
   (a) and its close-up view up to \rew\ = 1 \AA\ (b).  The
   corresponding plots for low-ion absorption lines are shown in lower
   panels (c,d).  The meaning of symbol colors are same as
   Figure~\ref{f1}. For 1on absorbers, only upper limits are shown
   with arrows. If data points locate on the dashed lines, the
   observed \rew\ in one sightline is equal to that in another
   sightline.\label{f2}}
\end{figure*}

Figure~\ref{f1} shows distributions of the physical distance in the
transverse direction (\dtra) between the sightlines for high-ion and
low-ion ionization absorbers as a function of absorption
redshift. Thick black curve denotes a physical distance between the
sightlines corresponding to the typical lensed quasar from SQLS
(\zem\ = 2.3, \zl\ = 0.5, and $\theta$ =
2.$^{\!\!\prime\prime}$0). For high-ions, our Gemini/GMOS and
Keck/ESI spectra sample absorbers with \dtra\ $\sim$ 0.1 -- 1~kpc at
\zabs\ $\sim$ 1.5 -- 2.5. \citet{ell04} sample absorbers at higher
redshift up to \zabs\ $\sim$ 3.6, while \citet{mon98} and
\citet{sme95} sample those with larger physical distance up to
\dtra\ $\sim$ 10~kpc.  On the other hand, for low-ions all data source
sample absorbers with \dtra\ $\sim$ 0.1 -- 10~kpc at \zabs\ $\sim$ 0.5
-- 2.0, although \citet{ell04} have several sample at higher redshift
up to \zabs\ $\sim$ 3. By combining our high ion samples (black and
red open circles in Figure~\ref{f1}) with that of \citet{ell04}
(blue open circles in Figure~\ref{f1}), we can examine the redshift
evolution of absorbers with a scale of \dtra\ $\sim$ 0.1 - 1~kpc. On
the other hand, we can examine physical properties of absorbers with a
wide range of distance in \dtra\ $\sim$ 0.1 - 10~kpc at \zabs\ $\sim$
2.

We first compare strengths of absorption lines (i.e., \rew) in the two
sightlines as shown in Figure~\ref{f2} for high and low-ions.  Samples
from \citet{ell04} and \citet{sme95} tend to have smaller \rews, while
those from \citet{mon98} and our sample have larger values.  This is
because the former are detected in spectra with higher S/N ratio
(i.e., due to a technical reason).  Correlation coefficients between
\rews\ along sightline pairs are almost same; $r$ = 0.981 and $r$ =
0.933 for high-ions and low-ions, respectively (see
Figure~\ref{f2}~(a) and (c)).  However, we find larger scatter for
low-ions ($r$ = 0.594) compared to high-ions ($r$ = 0.941) if we
consider only weak absorption lines with \rew\ smaller than 1.0 \AA, 
as noted in \citet{ell04} (see Figure~\ref{f2}~(b) and (d)). This could
be due to a clustering effect. If strong/weak absorption lines
correspond to regions with high/low number density of gas clouds, only
the weaker ones with a sparse cloud distribution are affected by a
typical scale of each cloud that should be smaller for low-ionized
absorbers.

\begin{figure}
 \centering
 \includegraphics[width=\columnwidth]{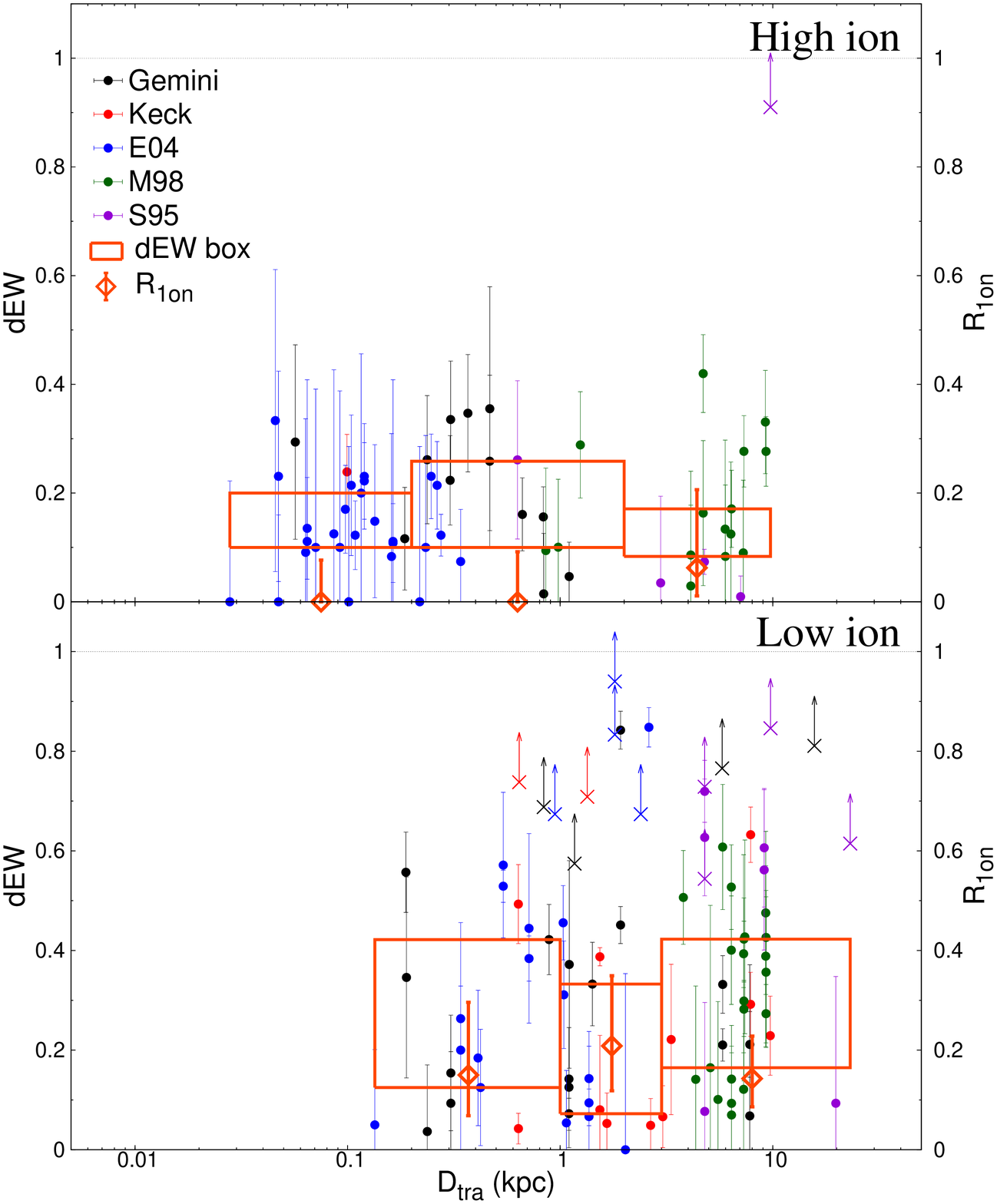}
 \caption{Fractional equivalent width (\dew) and 1on ratio (\rone) for
   high (top) and low (bottom) ions as a function of sightline
   separation (\dtra).  The meaning of symbol colors are same as
   Figures~\ref{f1} and \ref{f2}. If absorption lines are 1on sample,
   only lower limits of \dew\ are plotted with upward arrows.  The
   range of 30 -- 70 percentile of \dew\ distribution (\dew\ box) and
   1on ratio (\rone) are shown with orange rectangles and open
   diamonds after separating \dtra\ range into three bins.\label{f3}}
\end{figure}

Thus, a typical scale of absorbers probably depends on their
ionization condition; those in higher ionization condition tend to
have larger size \citep[e.g.,][]{ste16}. Therefore, we investigate the
fractional equivalent width difference \dew\ as a function of
transverse distance between sightlines as shown in Figure~\ref{f3}.
Because most absorption systems in our sample are detected at redshift
higher than those of lensing galaxies (i.e., \zabs\ $>$ \zl), the
corresponding physical separation between sightlines becomes larger at
lower redshift. Therefore, low-ions such as \ion{Mg}{2} and
\ion{Fe}{2}, whose rest-frame wavelengths are larger than those of
high-ions like \ion{C}{4} and \ion{Si}{4}, are detected at lower
redshifts and we can perform multiple sightline spectroscopy only for
larger separation distances.

We then compare the \dew\ distributions as a function of \dtra\ after
separating \dtra\ into three bins for high- and low-ion samples in
such a way that each bin contains almost same number of absorption
lines.  Because we confirmed that \dew\ distribution is not Gaussian,
we regard the range between 30 percentile and 70 percentile of the
\dew\ distribution in each bin as a core distribution range of
\dew\ (\dew\ box, hereafter; see Figure~\ref{f3}). The \dew\ box for
high ions are 0.10 -- 0.20, 0.10 -- 0.26, and 0.08 -- 0.17 in
\dtra\ of 0.02 -- 0.2, 0.2 -- 2.0, and 2.0 -- 10.0~kpc, while the
\dew\ box for low ions are 0.13 -- 0.42, 0.07 -- 0.33, and 0.17 --
0.42 in \dtra\ of 0.1 -- 1.0, 1.0 -- 3.0, and 3.0 -- 20~kpc,
respectively. We confirm that the \dew\ distributions are almost
independent of \dtra\ for both of high and low ion samples
(\dew\ $\sim$ 0.2) although the low-ion sample has a slightly larger
variation.  We also do not find any remarkable redshift evolution of
the \dew\ distribution for high-ion absorbers in the range of
\dtra\ $\sim$ 0.1 -- 1~kpc, comparing our sample at \zabs\ $\sim$ 2
(black and red filled circles in Figure~\ref{f3}) and those from
\citet{ell04} at \zabs\ $\sim$ 3.3 (blue filled circles in
Figure~\ref{f3}).

We also compare the fraction of 1on lines (1on ratio, hereafter)
defined by
\begin{equation}
R_{\rm 1on} = \frac{N_{\rm 1on}}{N_{\rm 2on}+N_{\rm 1on}},
\label{eqn2}
\end{equation}
where $N_{\rm 1on}$ and $N_{\rm 2on}$ are numbers of 1on and 2on
lines.  Because the \rone\ value strongly depends on the quality of
spectra (i.e., detection limit), we calculate the ratio using only
reliable 1on lines; 3$\sigma$ detection limit on \rew\ in an
undetected sightline is smaller than 50\%\ of \rew\ in a detected
sightline (i.e., 6$\sigma$(\rew$_{\rm undet}$) $<$ \rew$_{\rm
  det}$)\footnote[6]{$\rew_{\rm det}$ is a rest-frame equivalent width
  in spectra of detected sightline, while $\sigma(\rew_{\rm undet})$
  is a $1\sigma$ detection limit on \rew\ in spectra of undetected
  sightline.} to find larger values for low-ions (\rone\ $\sim$ 0.16)
than high-ions (\rone\ $\sim$ 0.02).  This result suggests that a
typical scale of low-ion absorbers is smaller than those of high-ion
absorbers, which is consistent with the results from the past studies
\citep[e.g.,][]{ell04,ste16} and the correlation analysis for weak
absorption lines (Figure~\ref{f2}~(b) and (d)).

\section{DISCUSSION}
\begin{figure*}
 \centering
 \includegraphics[width=7.5cm]{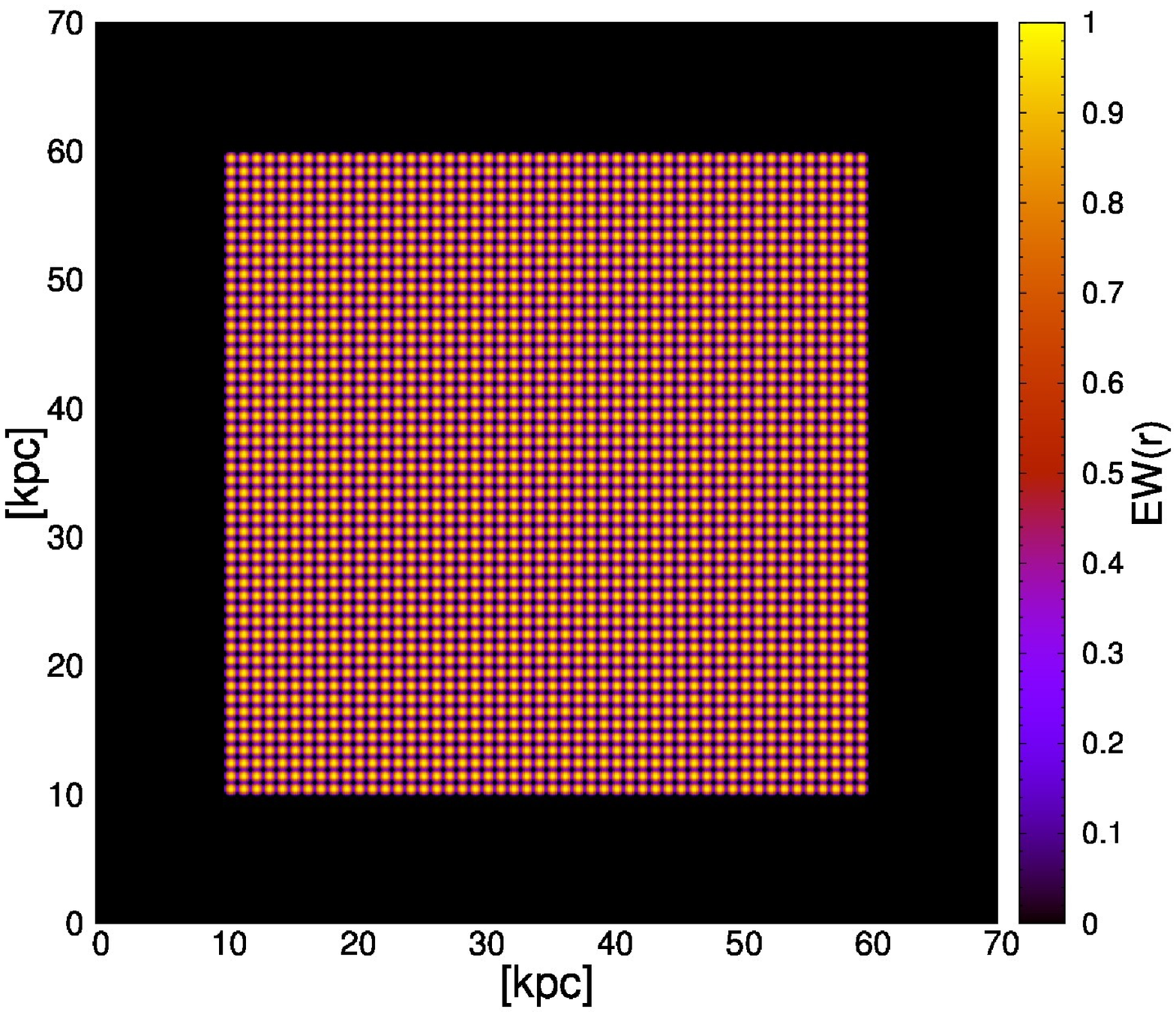}
 \includegraphics[width=7.5cm]{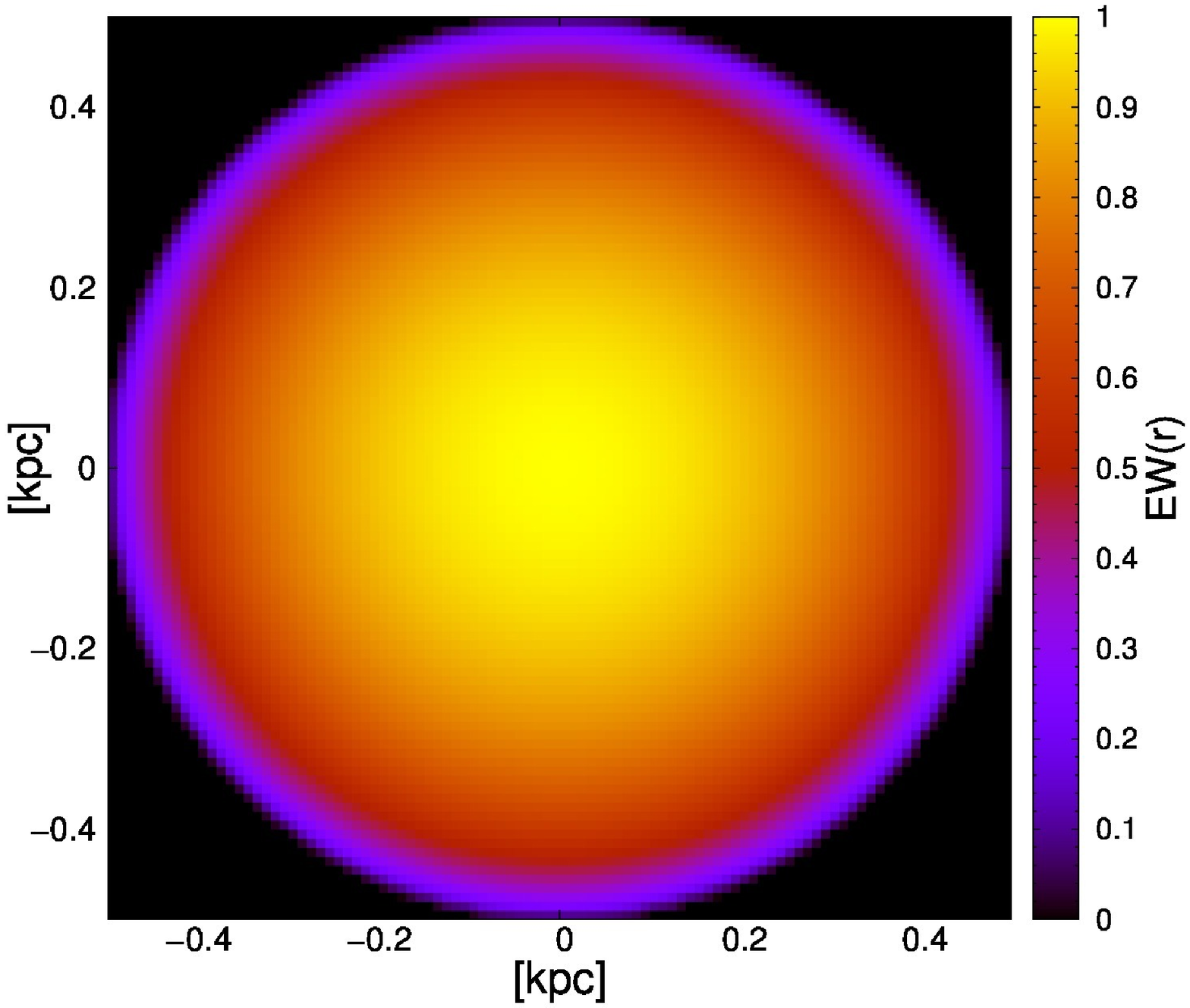}
 \vspace*{-15mm}
 \caption{Sample model of the CGM with a default parameters: an
   elliptical function, a size of each spherical cloud of $d$ = 1~kpc,
   an overall size of the CGM of $L$ = 50~kpc, a covering factor of
   \cf\ = 1 with no intensity of equivalent width by diffuse gas
   ($\ew_{\rm diff}$ = 0).  Left and right panels show a whole range of
   the modeled CGM and its close-up of a cloud.\label{f4}}
\end{figure*}

\begin{figure*}
 \centering
 \includegraphics[width=7cm]{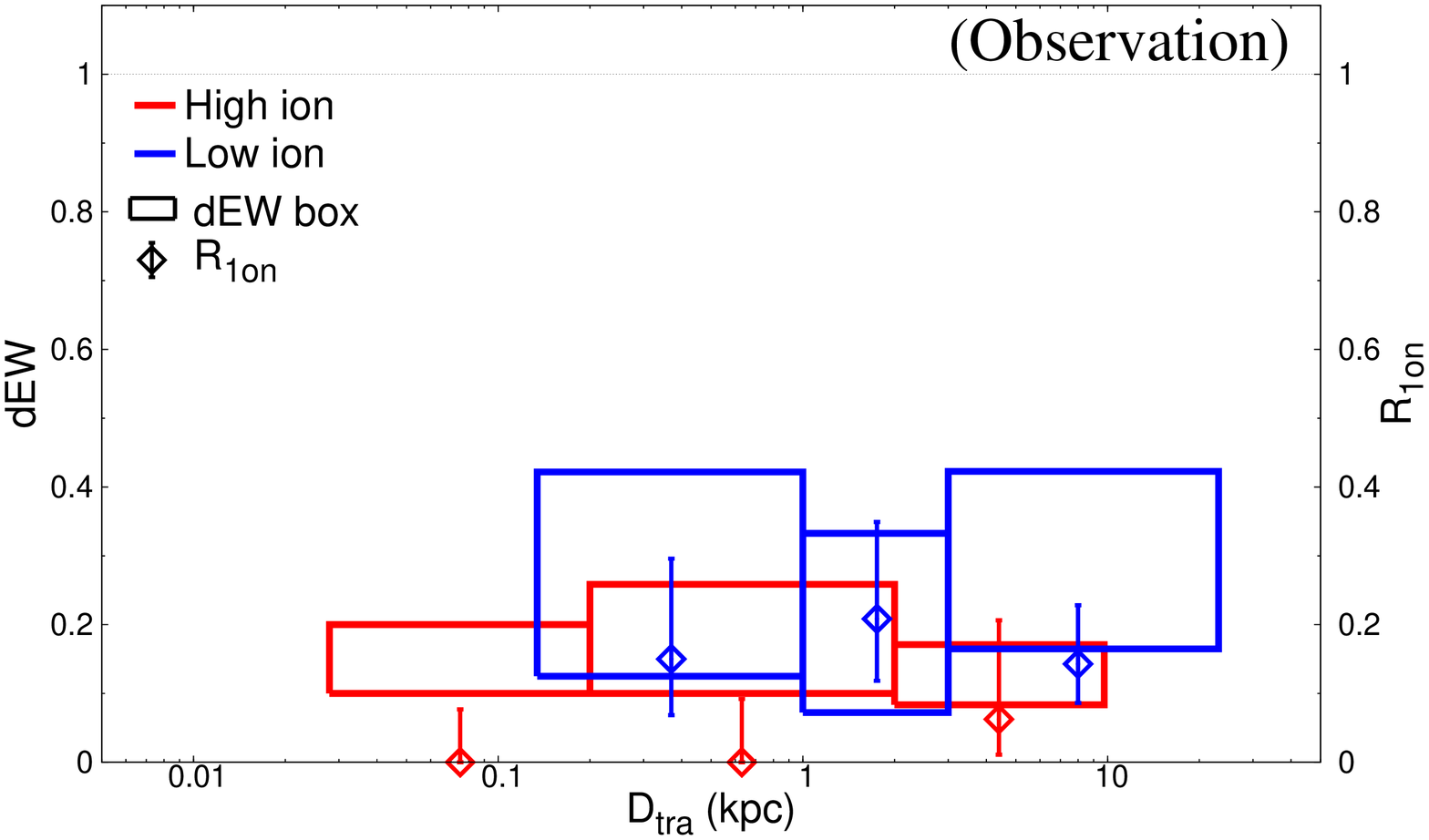}
 \includegraphics[width=7cm]{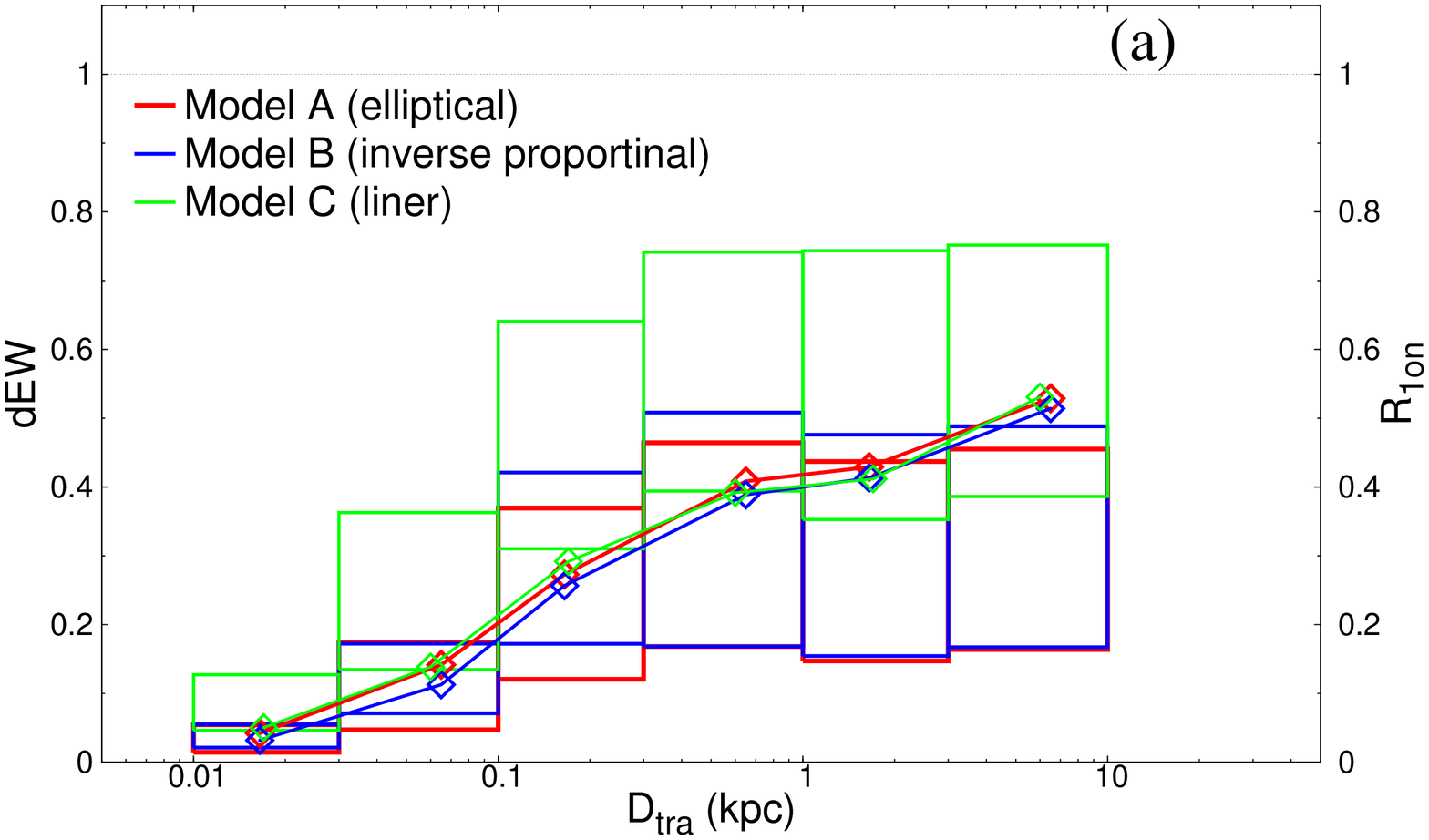}
 \includegraphics[width=7cm]{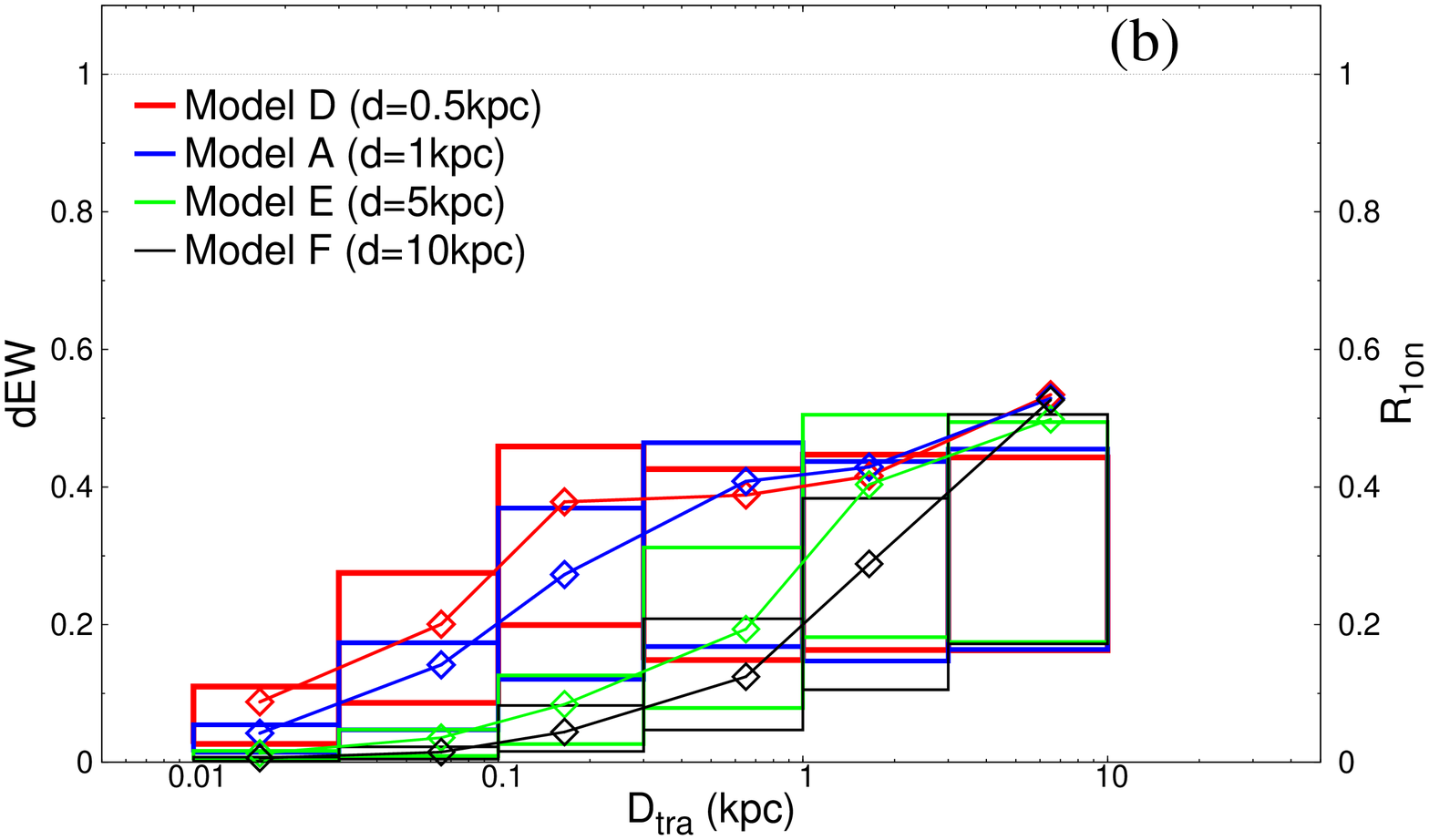}
 \includegraphics[width=7cm]{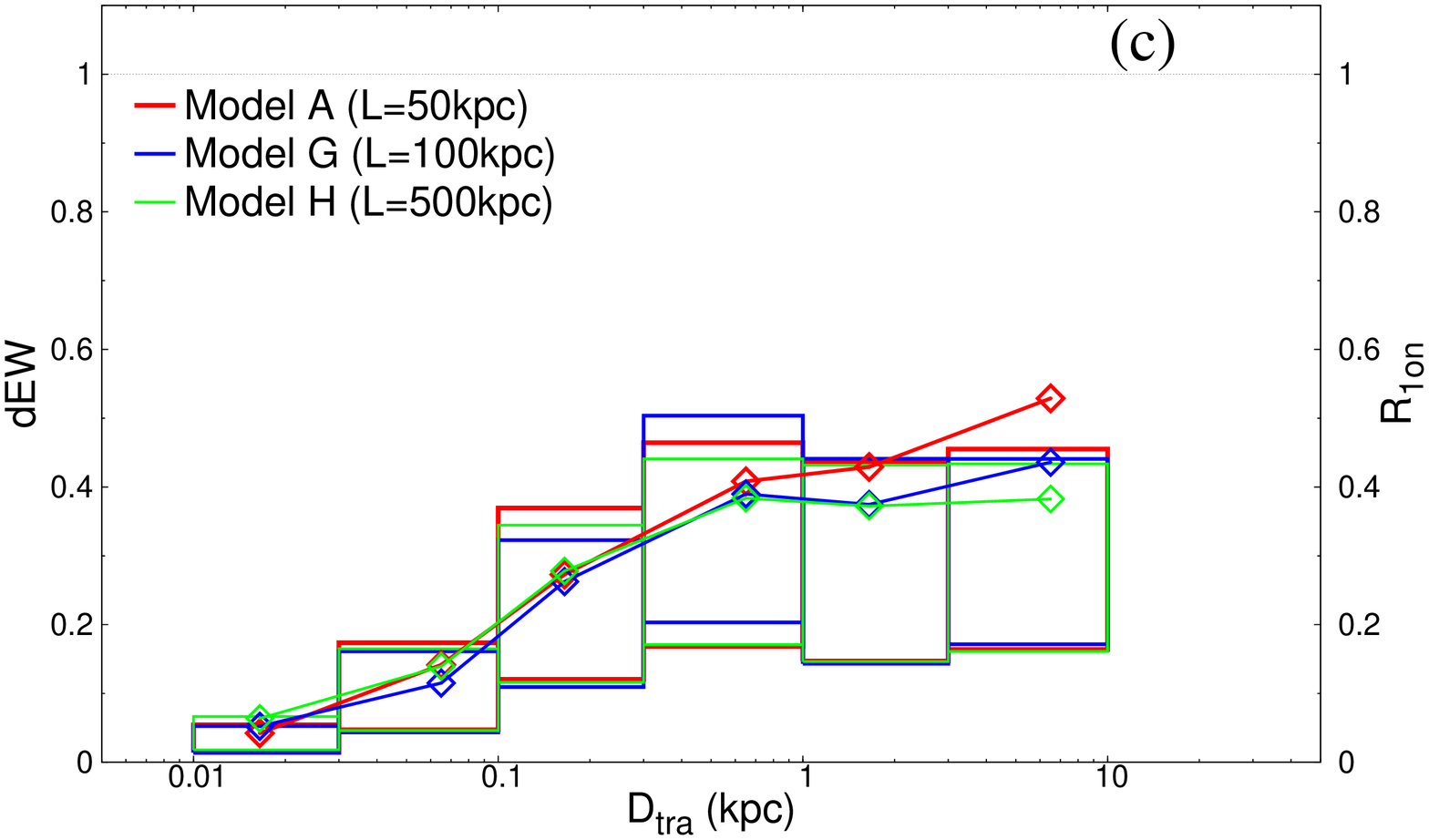}
 \includegraphics[width=7cm]{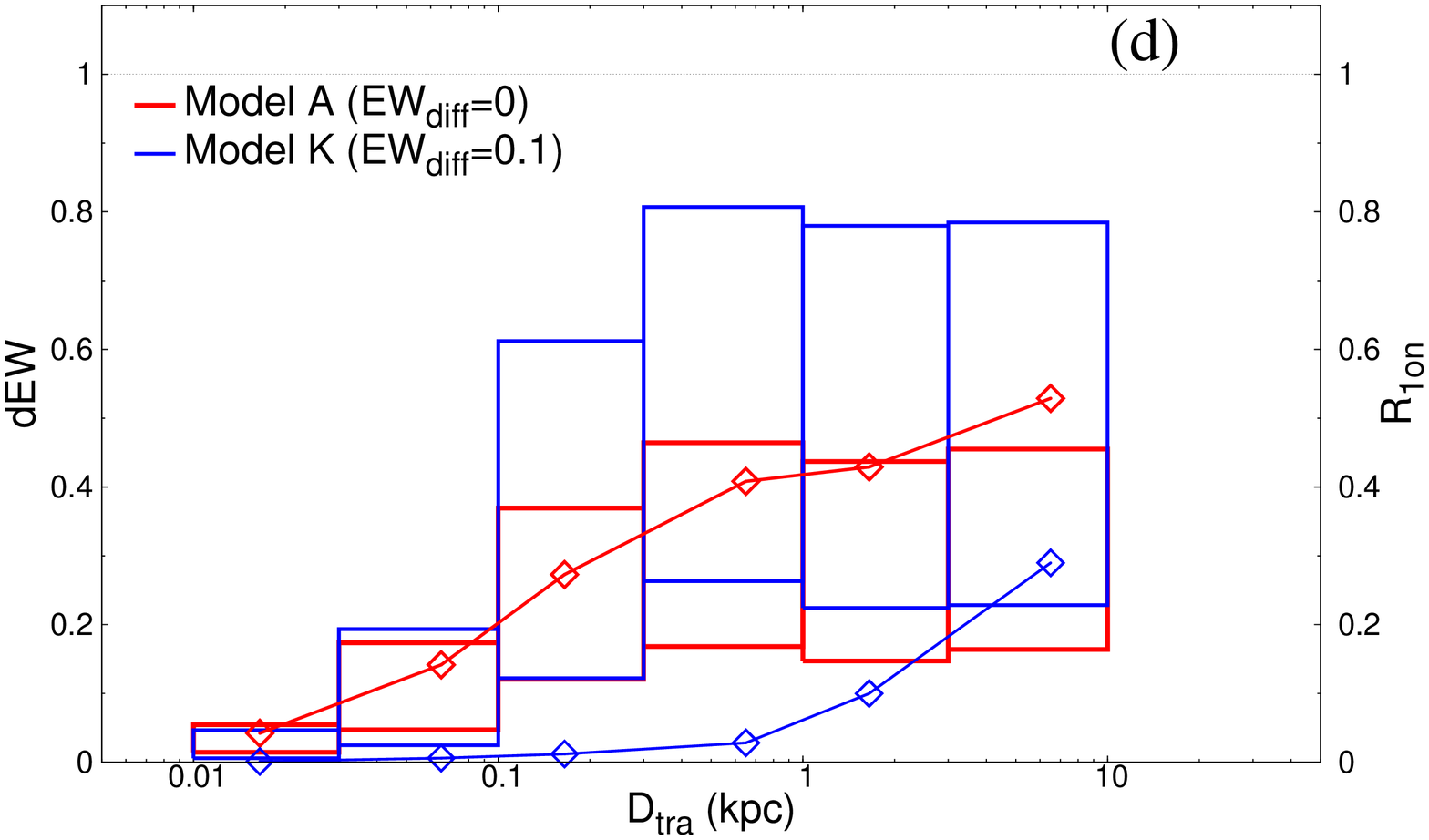}
 \includegraphics[width=7cm]{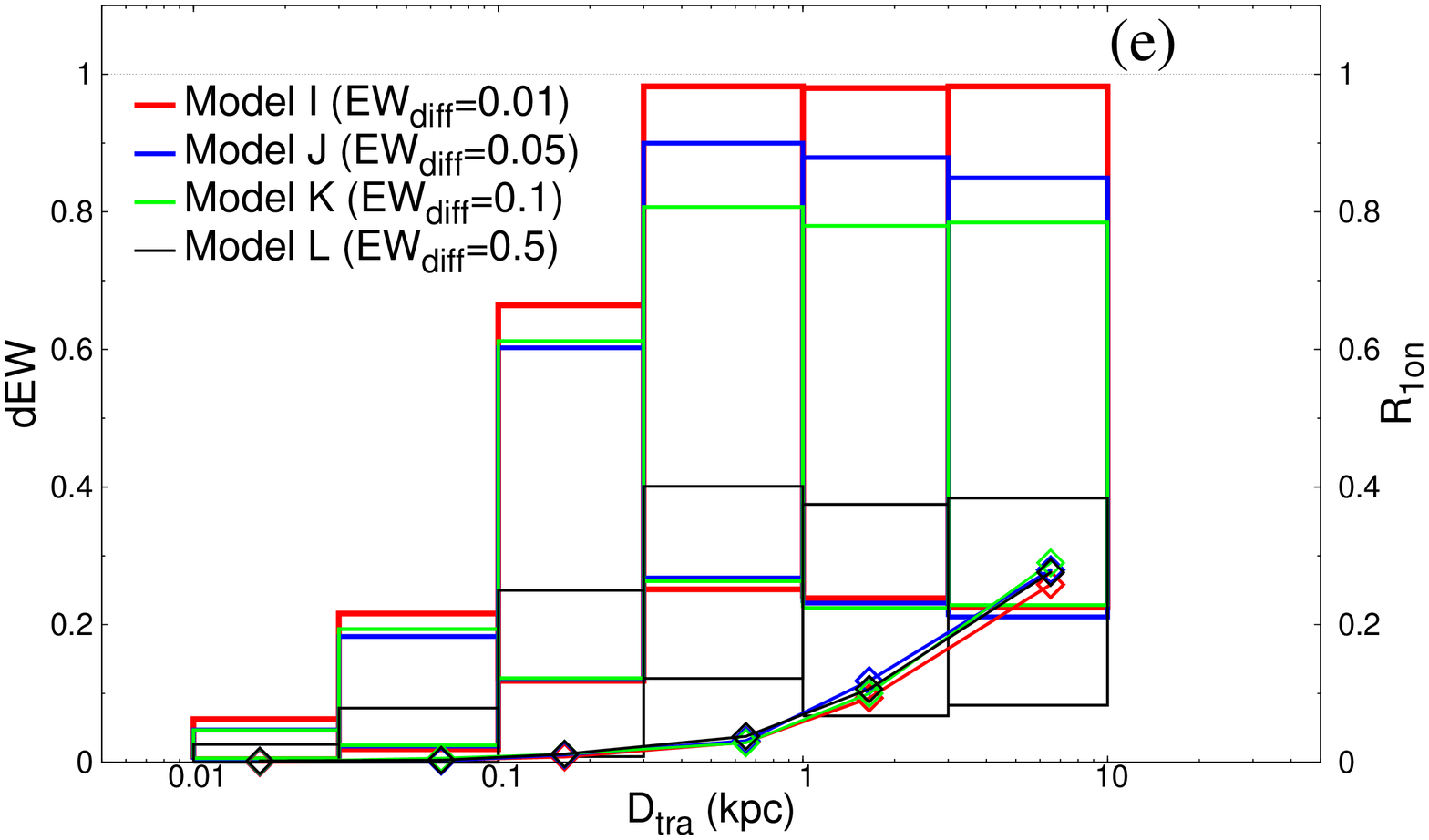}
 \includegraphics[width=7cm]{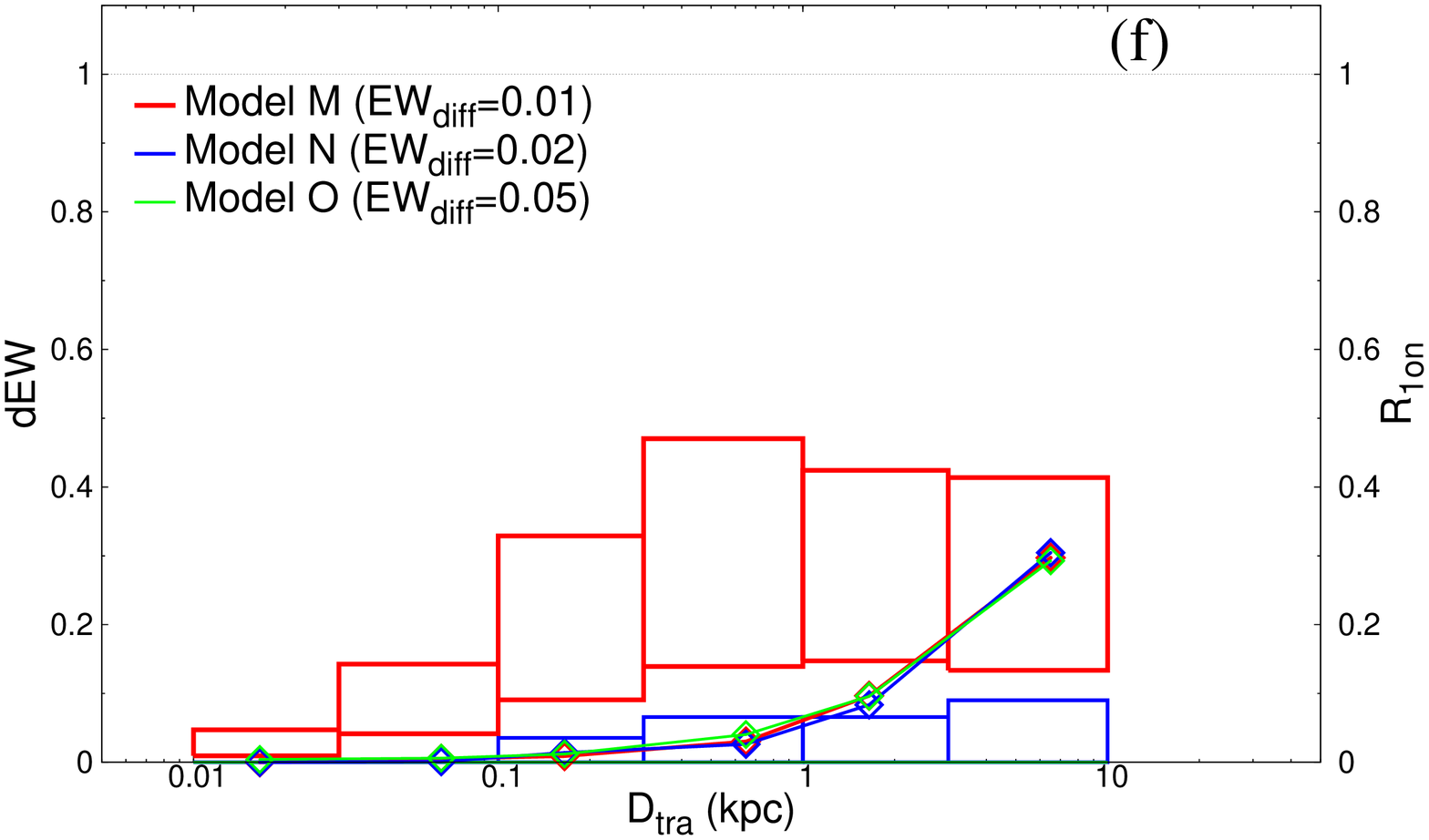}
 \includegraphics[width=7cm]{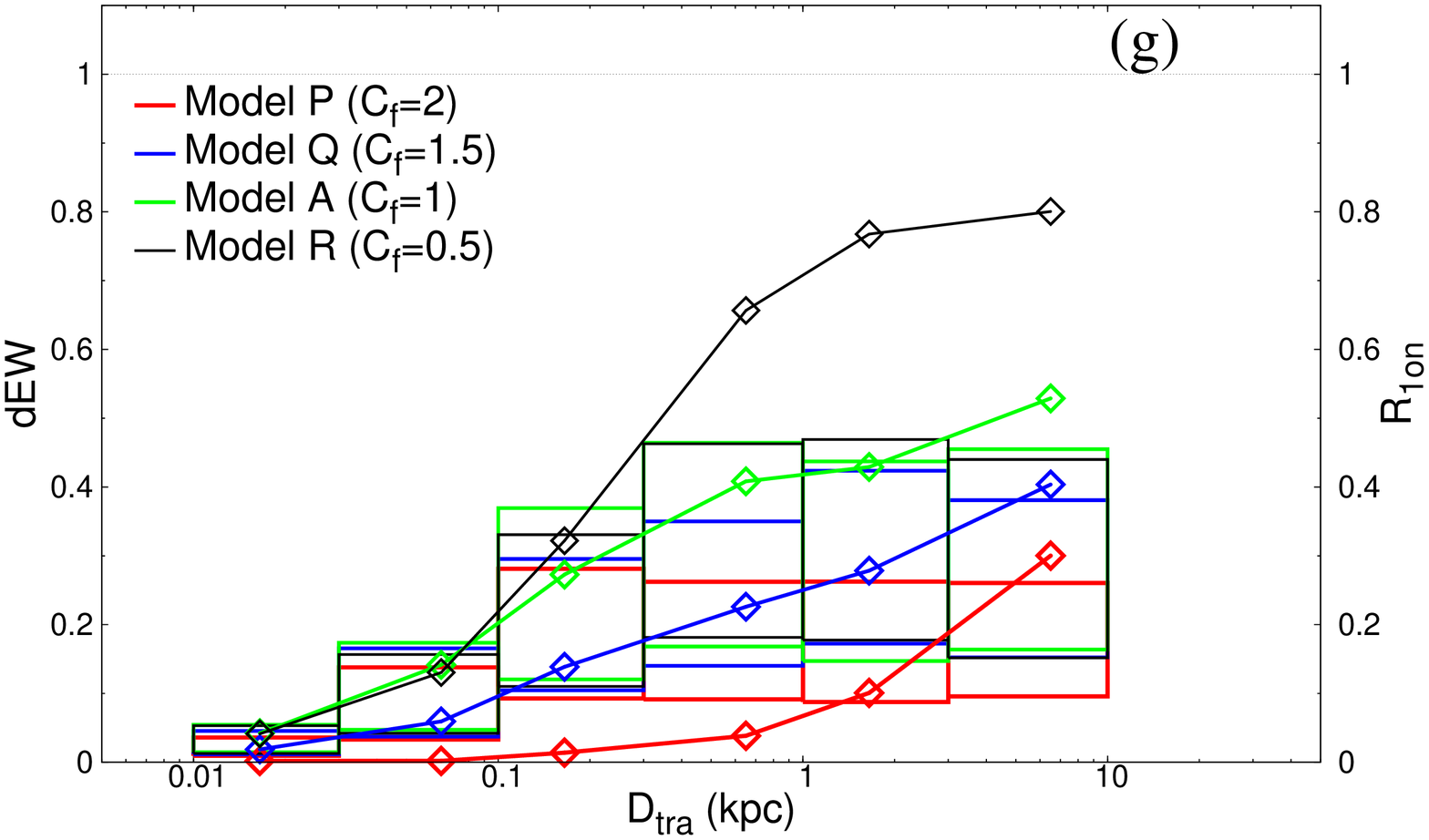}
 \caption{Same as Figure~\ref{f3}, but for models.  The range of 30 --
   70 percentile of \dew\ distribution and 1on ratio are shown with
   rectangles and open diamonds.  We adopt default condition (Model A
   in Table~\ref{t4}): spherical clouds with a size of $d$ = 1~kpc,
   whose radial distribution of equivalent width ($\ew(r)$) is written
   by an elliptical function, a covering factor of \cf\ = 1 with no
   intensity of equivalent width by diffuse gas ($\ew_{\rm diff}$ = 0)
   in a square region (i.e., CGM) with a length of one side of $L$ =
   50~kpc.  Results are shown for models if we change a function of
   radial distribution of equivalent width (a), a size of each
   spherical cloud (b), an overall size of the CGM (c), an addition of
   diffuse gas (d), a strength of diffuse gas for both functions
   (e,f), and a covering factor (g). See Table~\ref{t4} for adopted
   parameters in detail.\label{f5}}
\end{figure*}

In previous section, we discovered several properties on \dew\ and
\rone\ for high and low-ion absorbers. In order to connect our
findings with the internal structure of the CGM that cannot directly
be observed, we construct simple models to reproduce the observed
properties.  We assume a two-component model for the CGM; a number of
spherical gas clouds are embedded in diffusely distributed gas and
both of them give rise to the metal absorption lines.  Although the
ionization state depends on several parameters including electron
density, photon density, and gas temperature, we assume a single
  ionization state for each of the two-components, for simplicity. The
\ews\ of both components are summed to measure a total \ew, although
only spherical clouds have a radial gradient of \ew.  Thus, the total
equivalent width depends on a) an equivalent width distribution as a
function of radius $r$ from the center of each spherical cloud
(\ew($r$)) and b) an intensity of equivalent width by diffusely
distributed gas component ($\ew_{\rm diff}$). Focusing on the
probability that the spherical clouds and the diffuse component locate
along the two sightlines toward the background quasars, the cross
section of the absorbers depends on their sizes.  Here, we also define
c) a size (diameter) of each spherical cloud ($d$) and d) an overall
size of diffuse component ($L$).  Here, for simplicity, we assume that
a number of spherical clouds randomly distribute on scale of $L$ in a
square region\footnote[7]{Because models with clouds in {\it square}
  area and {\it circle} area give almost same results with only a few
  \%\ difference in \dew\ and \rone\ distributions, we adopt the
  former for our calculation.}.  The covering factor of the absorbing
clouds also determines the incidence rate of the clouds.  In this
model, we define e) a covering factor of clouds (\cf).  Using the five
parameters above (i.e., \ew($r$), $\ew_{\rm diff}$, $d$, $L$, and
\cf), we examine the incidence rate of gas clouds and the total
equivalent width in the two sightlines toward the background quasars.
Because high and low-ion absorption lines are not necessarily arising
at the same gas, we make models for each of them, respectively.  In
the above model, we use the equivalent width distribution instead of
the column density distribution because equivalent widths can be
compared to the observation directly.\footnote[8]{If absorption lines
  are not saturated (i.e., its central optical depth is $\tau_0$ $\ll$
  1), their equivalent widths are almost proportional to the column
  density at the linear part of the curve of growth. We can apply this
  assumption for a substantial fraction of our sample.}  First, we
place a number of spherical clouds in a square field with a 10~kpc
margin (that correspond to the maximum separation distance of our
sample) around the square field (see left panel of Figure~\ref{f4}).
And then, we divide sightline distance into 6 bins, randomly choose
1000 sightline pairs for each bin (i.e., 6000 random sightline in
total), and measure equivalent widths for them ($\ew_1$ and $\ew_2$).
We repeat such measurements for each model by changing five parameters
above to find the best model to reproduce the observations (see
Figure~\ref{f5} and Table~\ref{t4}).

\begin{deluxetable}{cccccc}
\tabletypesize{\tiny}
\tablecaption{Model Parameters\label{t4}}
\tablewidth{0pt}
\tablehead{
\colhead{Model}             &
\colhead{function$^a$}      &
\colhead{$d$$^b$}           &
\colhead{$L$$^c$}           &
\colhead{$\ew_{\rm diff}$$^d$} &
\colhead{\cf$^e$}           \\
\colhead{}                  &
\colhead{}                  &
\colhead{(kpc)}             &
\colhead{(kpc)}             &
\colhead{}             &
\colhead{}                  \\
\colhead{(1)}               &
\colhead{(2)}               &
\colhead{(3)}               &
\colhead{(4)}               &
\colhead{(5)}               &
\colhead{(6)}
}
\startdata
A & ell. &  1   &  50 & ...  & 1    \\
B & inv. &  1   &  50 & ...  & 1    \\
C & lin. &  1 &  50 & ...  & 1    \\
\hline
D & ell. &  0.5 &  50 & ...  & 1    \\
E & ell. &  5   &  50 & ...  & 1    \\
F & ell. & 10   &  50 & ...  & 1    \\
\hline
G & ell. &  1   & 100 & ...  & 1    \\
H & ell. &  1   & 500 & ...  & 1    \\
\hline
I & ell. &  1   &  50 & 0.01 & 1    \\
J & ell. &  1   &  50 & 0.05 & 1    \\
K & ell. &  1   &  50 & 0.1  & 1    \\
L & ell. &  1   &  50 & 0.5  & 1    \\
\hline
M & inv. &  1   &  50 & 0.01 & 1    \\
N & inv. &  1   &  50 & 0.02 & 1    \\
O & inv. &  1   &  50 & 0.05 & 1    \\
\hline
P & ell. &  1   &  50 & ...  & 2  \\
Q & ell. &  1   &  50 & ...  & 1.5  \\
R & ell. &  1   &  50 & ...  & 0.5  \\
\hline
Best Model 1 (high-ion) & ell. & 1   & 500 & ... & 2 \\
Best Model 2 (high-ion) & ell. & 0.5 & 500 & 0.5 & 1   \\
Best Model 3 (low-ion)  & ell. & 1   & 500 & ... & 1.5 
\enddata 
\tablenotetext{a}{Function of radial distribution of equivalent width:
  elliptical (ell.), \\ inverse proportional (inv.), and liner (lin.) functions.}
\tablenotetext{b}{Size of each absorbing cloud.}
\tablenotetext{c}{Overall size of the CGM.}
\tablenotetext{d}{Intensity of equivalent width in a diffuse gas. \\ Three dots means no diffuse gas is added.}
\tablenotetext{e}{Covering factor of clouds in the CGM.}
\end{deluxetable}

Among several free parameters, we first consider physical acceptable
functions for the radial distribution of the equivalent width. In this
paper, we assume three simple \ew($r$) functions described below.
\begin{description}
 \item[(a) Elliptical function] \mbox{} \\ If absorbers have a
   spherical shape with no internal structure (i.e., homogeneous
   density), an equivalent width is proportional to a projected depth
   of the absorber unless an absorbing cloud is optically thick. In
   this case, the equivalent width distribution is expressed by
   \begin{equation}
   \ew(r) = \ew_{\rm max}\sqrt{1-\frac{r^2}{(d/2)^2}},
   \end{equation}
   where $r$ is the distance from the center of each spherical
   cloud, $d$ is a diameter of the cloud, and $\ew_{\rm max}$ is an
   intensity of equivalent width at $r$ = 0. 
 \item[(b) Inverse proportional function] \mbox{} \\ Another possible
   model is a singular isothermal sphere with the radial density
   distribution of $\rho$($r$) $\propto$ $r^{-2}$. In this model, a
   projected density (i.e., column density) at a distance from the
   center $r$ is approximately expressed by an inverse proportional
   function except for at very large radius. To avoid it from diverge
   to infinity at the center, we slightly change the function into
   \begin{equation}
   \ew(r) = 
   \frac{
   \ew_{\rm max} \left(\frac{\ew_{\rm min}}{\ew_{\rm max}-\ew_{\rm
       min}}\frac{d}{2}\right) 
       }
       { \left(r+\frac{\ew_{\rm min}}{\ew_{\rm
       max}-\ew_{\rm min}}\frac{d}{2}\right)},
   \end{equation}
   where $\ew_{\rm min}$ is the minimum observational value of
     \ew\ in our sample.
 \item[(c) Linear function] \mbox{} \\ For comparison to the results from 
 models adopting the above equivalent width distributions, we also examine 
 a simple model expressed by
   \begin{equation}
   \ew(r) = -\frac{\ew_{\rm max}}{(d/2)} r + \ew_{\rm max}.
   \end{equation}
\end{description}

We adopt a dimensionless number to set $\ew_{\rm max}$ = 1 as the
maximum equivalent width at the center, because we only measure the
fractional equivalent width difference (\dew) (i.e., the amplitudes of
$\ew_{\rm max}$ and $\ew_{\rm min}$ themselves do not necessarily have
to be measured.).  We also regard sightlines with equivalent widths
greater than $\ew_{\rm min}$ = 0.01 as absorption-detected sightlines
because the ratio of the maximum and the minimum \rews\ in our
observed sample is $\sim$100. The radial distribution functions for an
equivalent width depend mainly on the \dew\ distribution as a function
of \dtra. As shown in Figure~\ref{f5}~(a), the elliptical and the
inverse proportional functions show almost same patterns of \dew\ and
\rone\ distributions as a function of \dtra\ that well match to the
observed trends (i.e., the \dew\ box is $\sim$ 0.2 and almost
independent of \dtra.)  although \rone\ is rather
over-estimated. Among the two acceptable functions, we will use the
elliptical function as our default model. As for the other parameters,
we use $d$ = 1~kpc\footnote[9]{This is a typical size of \ion{N}{2}
  absorbers whose ionization parameter IP = 29.6~eV is between those
  of \ion{C}{4} (64.5~eV) and \ion{Mg}{2} (15.0~eV) \citep{ste16}},
$L$ = 50~kpc\footnote[10]{This is large enough (five times larger)
  compare to the maximum scale of our observation, $\sim$ 10~kpc}, and
\cf\ = 1 with no diffusely distributed homogeneous gas ($\ew_{\rm diff}
= 0$) as default parameters.

Next, we consider models with different sizes of each spherical cloud,
$d$ = 0.5, 1, 5, and 10~kpc using defaults values for all the other
parameters. It is clear that both \dew\ and \rone\ start to rise at
smaller \dtra\ for models with smaller cloud size (see
Figure~\ref{f5}~(b)).  We also change an overall size $L$ from 50,
100, to 500~kpc. As shown in Figure~\ref{f5}~(c), any clear
differences are not seen in the \dew\ distribution.  On the other
hand, \rone\ tends to have larger values at \dtra\ $\geq$ 1~kpc
especially for models with smaller overall size.  This is because a
number of 1on pairs that locate at the edge of overall area increase
for those models.

In addition to the clumpy spherical gas clouds, the CGM could have a
diffusely distributed homogeneous gas whose ionization condition
  is higher than that of the spherical clouds because of lower gas
  density. We add such a component (we call this {\it diffuse gas},
hereafter) whose equivalent width is 10\%\ of the central value (i.e.,
$\ew_{\rm diff}$ = 0.1) throughout the overall area.  Once the diffuse
gas is added, 1on pairs are seen only at the edge of overall area,
which decreases \rone\ significantly as shown in
Figure~\ref{f5}~(d). We also consider models with four diffuse gas
intensities with $\ew_{\rm diff}$ = 0.01, 0.05, 0.1, and 0.5 for the
elliptical function and $\ew_{\rm diff}$ = 0.01, 0.02, and 0.05 for
the inverse proportional function.  Although both functions have
acceptable models, their diffuse gas intensities are very different;
best models have diffuse gas intensity of $\ew_{\rm diff}$ = 0.5 and
0.01 for the elliptical and the inverse proportional functions,
respectively (Figure~\ref{f5}~(e) and (f)).

Lastly, we consider the covering factor \cf, which is the fraction of
the overall area that is covered by a number of spherical clouds, in
Figure~\ref{f5}~(g).  Here, we regard \cf\ = 1 if clumpy clouds are
regularly arranged with no gaps between clouds although there are
cracks\footnote[11]{Gaps between spherical clouds seen from us, toward
  which the corresponding equivalent width is zero.}  in the diagonal
directions (see Figure~\ref{f4}). If we decrease the covering factor
down to 0.5, \rone\ increases significantly to become inconsistent
with the observation (Figure~\ref{f5}~(g)). Therefore, we also attempt
to increase \cf\ to 1.5 and 2 that means 50\%\ or 100\%\ of the cracks
are covered by other foreground/background clouds along our
sightlines.  As show in Figure~\ref{f5}~(d) and (g), the
\rone\ distributions of the model with the \cf\ = 2 (Model P in
Table~\ref{t4}) and the model with the diffuse gas (e.g., Model K in
Table~\ref{t4}) are very similar to each other.  Because there are no
cracks inside the CGM in both the models, 1on sample occurs only at
the edge of the modeled CGM.

\subsection{Optimized Models}

\begin{figure}
 \centering
 \includegraphics[width=\columnwidth]{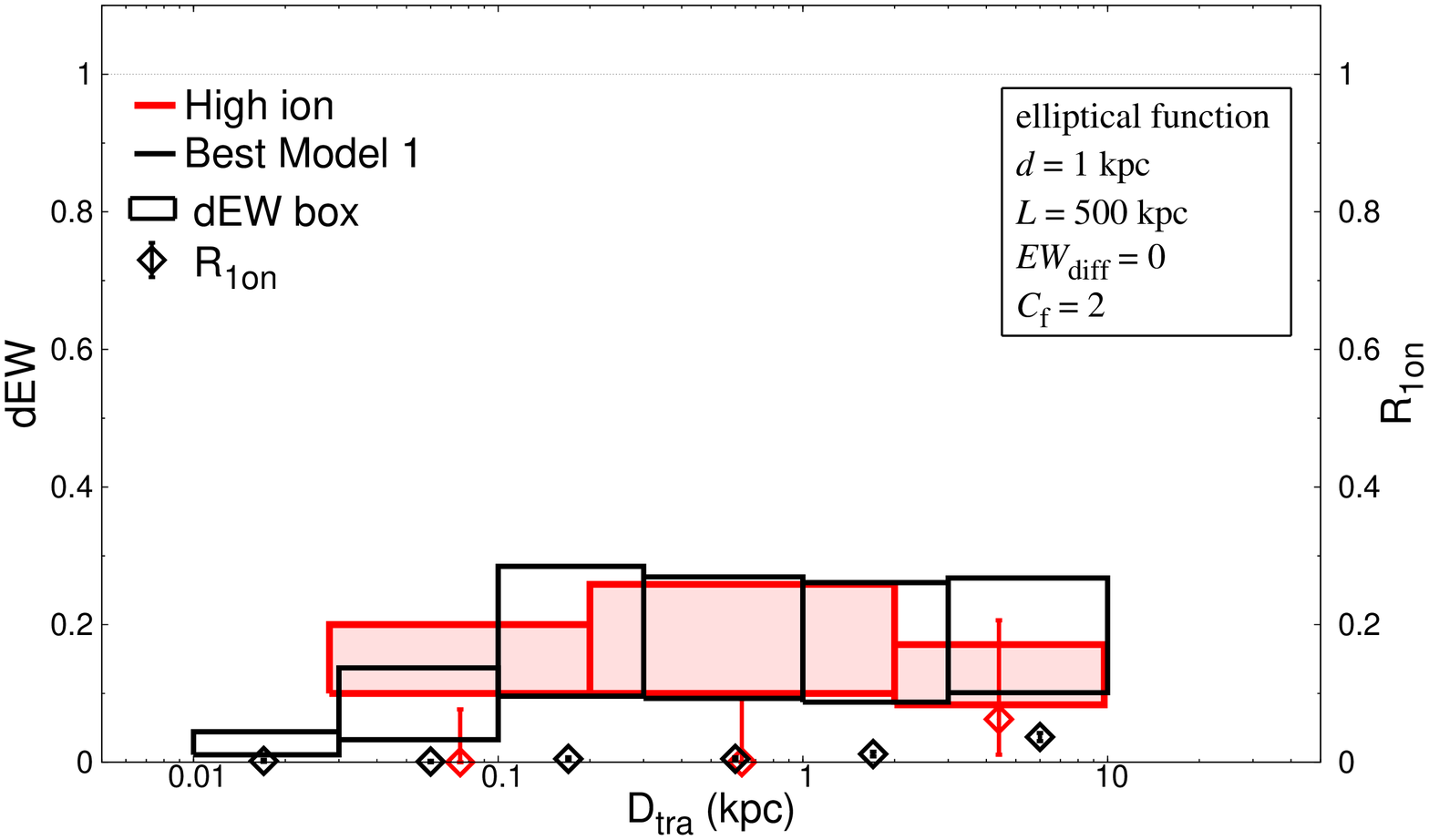}
 \includegraphics[width=\columnwidth]{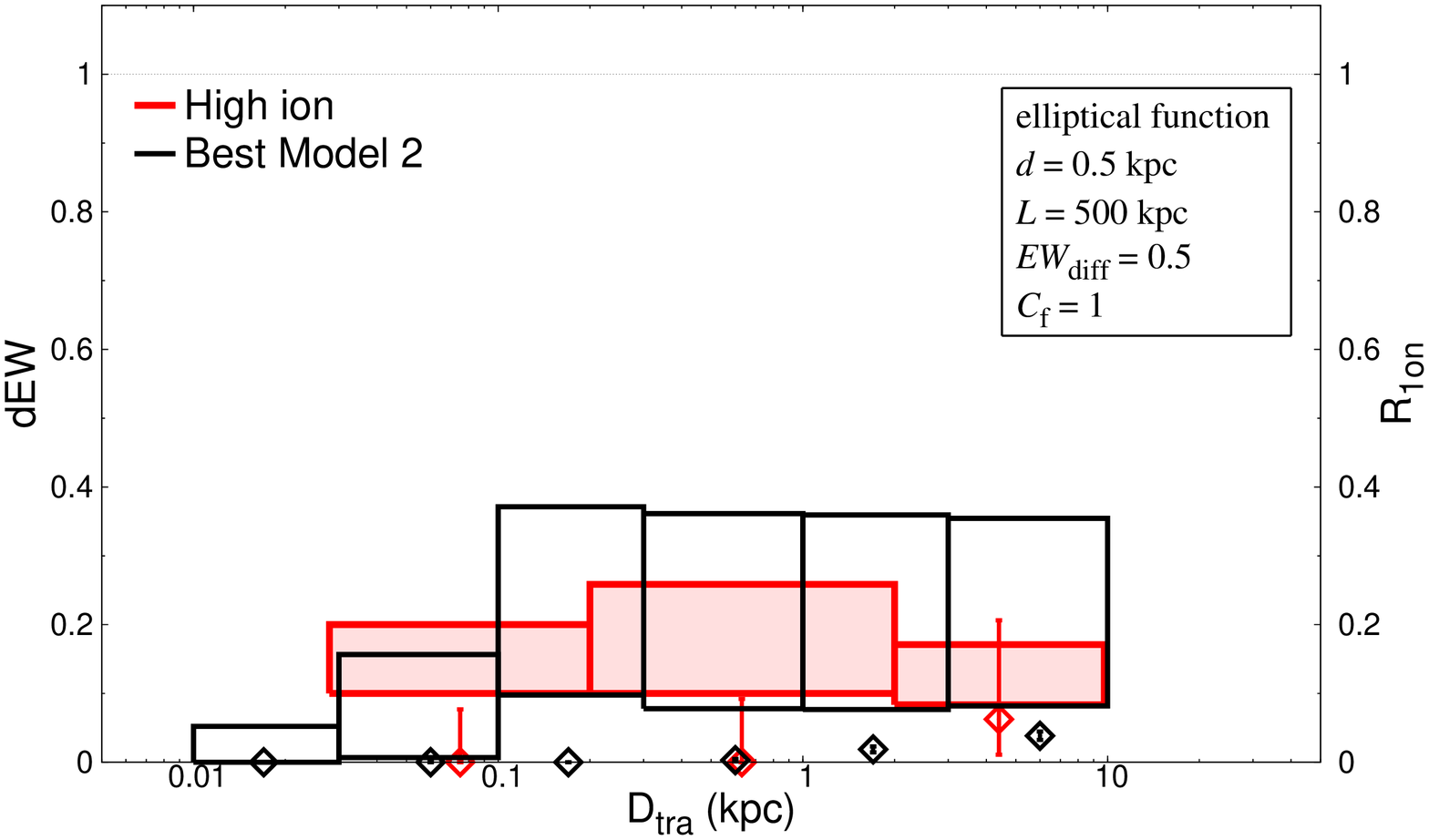}
 \includegraphics[width=\columnwidth]{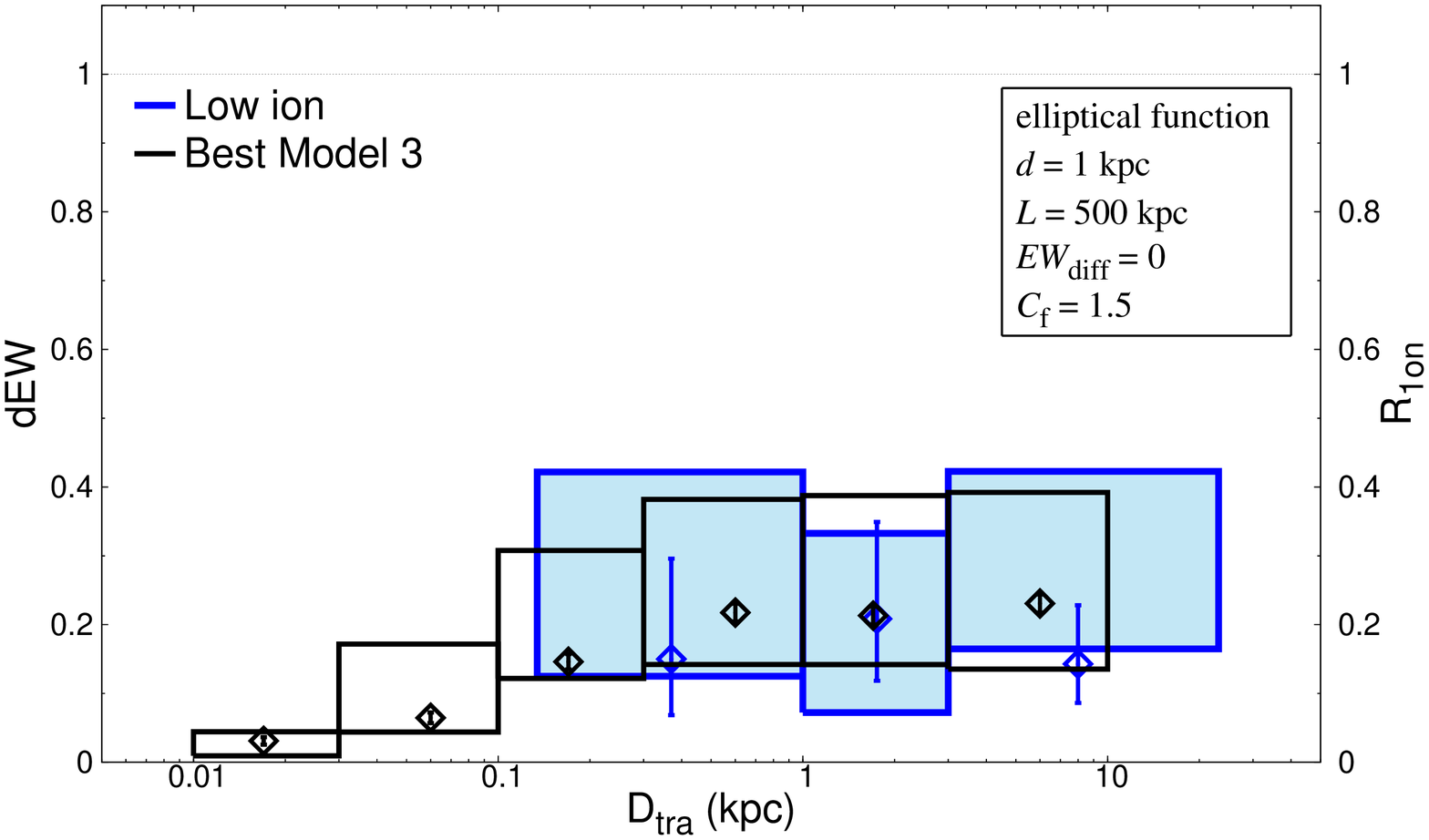}
 \caption{Same as Figure~\ref{f5}, but for the best models for
   high-ions (top and middle) and low-ions (bottom).  A colored
   rectangles and open diamonds indicates observed values, while a
   black ones indicate the model values. The best parameters for
   high-ions are elliptical function, $d$ = 1~kpc, $L$ = 500~kpc,
   \cf\ = 2, with no intensity of \ew\ by diffuse gas (top panel) or
   elliptical function, $d$ = 0.5~kpc, $L$ = 500~kpc, \cf\ = 1, with
   an intensity of $\ew_{\rm diff}$ = 0.5 (middle panel).  The best
   parameters for low-ions are elliptical function, $d$ = 1~kpc, $L$ =
   500~kpc, \cf\ = 1.5, with no diffuse gas (bottom panel).\label{f6}}
\end{figure}

After repeating above models by changing five parameters, we select
the best models for high and low-ion samples respectively, as shown in
Figure~\ref{f6} whose best parameters are summarized in
Table~\ref{t4}.  We chose the best models based on the following two
criteria; 1) \rone\ values from the model and the observation are
consistent to each other within 1$\sigma$ errors for a wide range of
\dtra, and 2) the model and the observation have the largest common
areas of \dew\ boxes among those satisfying the first criterion.  As
summarized below, our best models suggest both high and low-ion
absorbers have large (or full) coverage fraction along our sightlines.
However, the coverage fraction (i.e., a volume number density of gas
clouds) should be inversely proportional to the distance from host
galaxies \citep[e.g.,][]{rud12}, while we assume a constant cloud
density throughout the CGM. Our observation can be biased for regions
in the vicinity of host galaxies (e.g., $\leq$ 100~kpc) where the
coverage fraction is probably very high.

\begin{figure}
 \centering
 \includegraphics[width=\columnwidth]{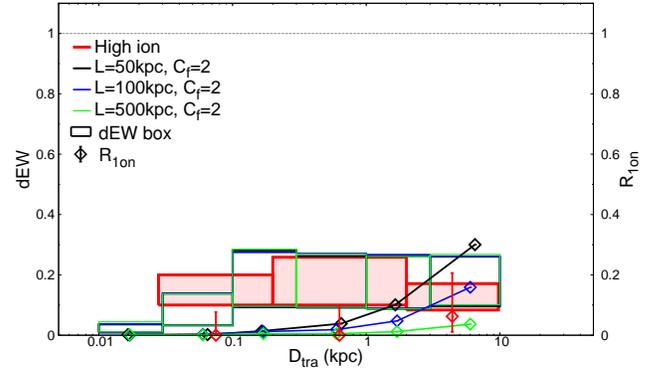}
 \caption{Same as Figure~\ref{f5}~(c), but using the best value for a
   covering factor (\cf\ = 2) instead of the default value (\cf\ = 1).
   The observed results for high-ions are also overlaid.\label{f7}}
\end{figure}

First, we focus on high ion absorbers. As shown in Figure~\ref{f3}, an
average value of the fractional difference of equivalent widths (\dew)
is obviously larger than $\sim$ 0.1 at the scale of \dtra\ $\sim$
0.1~kpc, which means that there exist small-size clouds or density
fluctuations at the corresponding scale.  As shown in
Figure~\ref{f5}~(b), the size of each cloud should be smaller than
$\sim$ 1~kpc. Otherwise, \dew\ is under-estimated at \dtra\ $\sim$
0.1~kpc. The 1on ratio (\rone) leaves zero only at larger scales of
several kilo-parsecs, which requires us to choose an overall size of
the CGM larger than $L$ $\sim$ 100~kpc (see Figure~\ref{f7}) once we
adopt the best value for a covering factor (\cf\ = 2) later.  It is
probably comparable to or larger than the overall size for low-ion
absorber ($L$ $>$ 500~kpc as described below) because high-ionized
absorbers usually have a larger distribution than low-ionized
absorbers \citep[e.g.,][]{ste16}.  High-ion absorbers cannot have any
small-scale cracks therein because their \rone\ is zero at
\dtra\ smaller than several kilo-parsec. Based on the above
considerations, we find two best models for high-ion absorbers.  They
have an overall size of $\sim$ 500~kpc or more with clumpy clouds (or
density fluctuation) smaller than 1~kpc whose equivalent widths
distribution follows the elliptical function\footnote[12]{The model
  using the inverse proportional function is also acceptable, but the
  elliptical function gives a better result.}. They have no cracks
between small clouds, which can be reproduced by either \cf\ = 2 or a
diffuse gas with an intensity of $\ew_{\rm diff}$ = 0.5
(Figure~\ref{f6} and Table~\ref{t4}).  Here, we emphasize that the
best size of each cloud is $d$ $\leq$ 0.5~kpc (instead of $\leq$
1~kpc) in the latter case.  Thus, we infer that high-ion absorbers
originate in a widely distributed homogeneous gas with a scale of
$\geq$ 500~kpc, in which there exists a small scale ($\leq$ 0.5 --
1~kpc) fluctuation of equivalent width.  Compared to our best model,
the past results for the CGM at $z$ $<$ 0.1 suggested they have a
smaller overall size of $\sim$ 100 -- 200~kpc around galaxies
\citep{che01,bor14,bur16}. As to covering factor, \citet{che01} got a
similar result to ours (i.e., full coverage) within $\leq$ 100~kpc
from host galaxies, although a partial coverage was sometimes
suggested even at smaller impact parameters
\citep[e.g.,][]{bor14}. The detection rate of \ion{C}{4} absorption
lines also depends on star-formation rate and stellar mass of host
galaxies as well as a Mpc scale galaxy number density around host
galaxies.

For low-ion absorbers, \dew\ is almost flat at \dtra\ $\sim$ 0.1 --
10~kpc like high-ion absorbers.  Such distributions can only be
reproduced by small clumpy clouds with scales smaller than $d$ $\leq$
1~kpc (see Figure~\ref{f5}~(b)). The 1on ratio is also flat at the
same \dtra\ range, which requires an overall size greater than $L$
$\sim$ 500~kpc. Covering factor (\cf) should be larger than 1 (to
avoid an over-estimation of \rone) but smaller than 2 (to make sub-kpc
scale gaps so that \rone\ is not zero). Our best model for low-ion
absorbers has an overall size of $\sim$ 500~kpc or more with small
clumpy clouds (or fluctuation) with a scale of $\sim$ 1~kpc. These are
same as those for high-ion absorbers. However, their covering factor
is smaller (\cf\ = 1.5) than high-ion absorbers (\cf\ = 2), which
means that low-ion absorbers have cracks with a scale of $\leq$ 1~kpc
(see Figure~\ref{f6} and Table~\ref{t4}). This is why only low-ion
absorbers have a substantial 1on ratio as small as \dtra\ $\sim$
0.1~kpc. For the same reason, low-ion absorbers do not have a
homogeneous gas component (i.e., $\ew_{\rm diff}$ = 0). Therefore, we
expect that a low-ion absorber consists of a large number of clumpy
dense clouds with a scale of $\leq$ 1~kpc.  Again, the past results
predicted a smaller size of the CGM ($\sim$ 200~kpc;
\citealt[e.g.,][]{nie13,chu13}) based on \ion{Mg}{2} absorption lines
at $z$ $<$ 1, compared to our best model.  The covering factor of
\ion{Mg}{2} was also estimated as \cf\ $\sim$ 0.6 -- 0.9, depending on
an azimuthal angle; the higher covering factors along the projected
galaxy major and minor axes \citep{kac12}.

\begin{figure}
 \centering
 \includegraphics[width=\columnwidth]{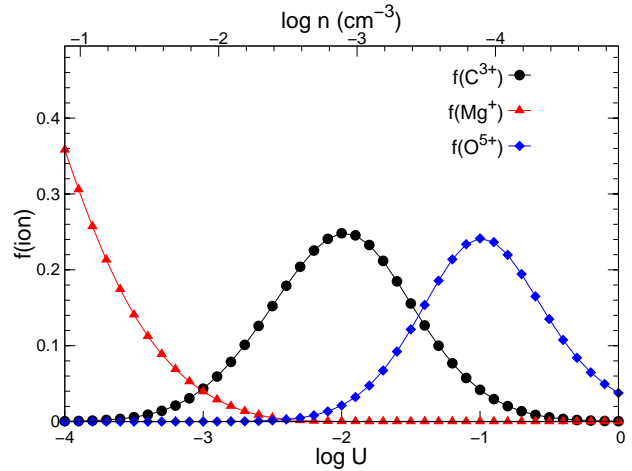}
 \caption{Ionization fractions ($f({\rm ion})$) of C$^{3+}$, Mg$^{+}$, and
     O$^{5+}$ as a function of ionization parameter ($\log U$) (bottom label) and
     a gas volume density ($\log n$) at $z$ = 2.0 (top label).  Photoionization
     models are conducted using the code Cloudy (ver. 17.00), assuming
     a plane-parallel slab that is illuminated with the extragalactic
     background radiation at $z$ = 2.0.  We also assume a constant
     metallicity of $\log(Z/Z_{\odot})$ = $-$1.0 throughout the gas
     whose total hydrogen column density is $\log N_{\rm H}$ =
     18.0~\cmm.\label{f8}}
\end{figure}

Thus, the best model above requires three components in total:
  i.e., clumpy clouds for high-ions, clumpy clouds for low-ions, and
  diffuse homogeneous gas for high-ions.  Obviously, these three
  components have different ionization conditions.  To locate the
  origins of them, we performed simple calculations using the
  photoionization code Cloudy, version 17.00 \citep{fer17}, on a slab
  of gas illuminated with extragalactic background
  radiation \citep{haa12} at $z$ = 2.0 (a typical \zabs\ in our
  sample).  We assume a constant metallicity of $\log(Z/Z_{\odot})$ =
  $-$1.0 throughout the gas whose total hydrogen column density is
  $\log N_{\rm H}$ = 18.0~\cmm\ in the optically thin
  regime\footnote[13]{At the range of ionization parameter that we
    modeled ($\log U$ = $-$4.0 -- 0.0), a total column density of
    neutral hydrogen is always smaller than $\log N_{\rm HI}$ =
    17.2~\cmm.}.  We varied the ionization parameter ($\log U$) in
  steps of 0.1~dex from $-$4.0 to 0.0 (which corresponds to the gas
  density of $\log n$ $\sim$ $-$0.6 -- $-$4.6 at $z$ = 2.0), and then
  calculated the ionization fractions of C$^{3+}$ (\ion{C}{4}),
  Mg$^{+}$ (\ion{Mg}{2}), and O$^{5+}$ (\ion{O}{6}) (as {\it an example} of
  higher ionized absorbers)\footnote[14]{In this assumption, since the gas temperature
   is $\sim$ 4 $\times$ $10^{5}$ ${\rm K}$, collision excitation does not 
   substantially contribute.}.  As shown in Figure~\ref{f8}, the ionization 
   fractions of Mg$^{+}$ and C$^{3+}$ are dominant at $\log
  U$ $<$ $-$3 ($\log n > -2$) and $\log U$ $\sim$ $-$3 -- $-$1 ($\log n \sim -2$ -- $-4$) 
  respectively, which are probably optimal for the \ion{C}{4} and \ion{Mg}{2} absorbers. 
  On the other hand, at $\log U$ $>$ $-$1 ($\log n < -4$), the fraction of O$^{5+}$
  becomes dominant although a substantial fraction of Carbon still
  remains in C$^{3+}$. Thus, the \ion{O}{6} absorbers in higher
  ionization condition that are frequently detected in the CGM
  \citep{tum11,tur14} could be the origin of the diffusely distributed
  \ion{C}{4} absorbers, while the spherical components of \ion{C}{4}
  and \ion{Mg}{2} absorbers correspond to the gas whose ionization
  parameters are optimal for them.

\subsection{Caveats}
Here, we list some caveats on our results that should be noted:

\begin{itemize}
\item{The source of absorber (i.e., a host galaxy) was not identified.
  There are several origins including dwarf galaxy, low surface
  brightness galaxy, galaxy merger, galaxy outflow, AGN outflow, and
  so on.  Without identifying host galaxies, we cannot narrow down
  possible sources.  Because of the same reason, we did not make a
  ``0on'' sample (i.e., no absorption is detected in both sightlines
  even if there exist a galaxy close to our sightlines to the
  quasars.) in addition to 1on and 2on samples because we do not know
  the locations of galaxies close to our sightlines in advance, which
  could over-estimate a covering factor (\cf\ $\geq$ 1.5) and also a
  size of the CGM ($L$ $\geq$ 500~kpc).}
\item{Only five free parameters are obviously not enough to make
  models. It can be improved by adding radial functions of gas density
  and volume density of spherical cloud from the center of the CGM as
  additional free parameters.  Indeed, both optical depth and
  equivalent width of \ion{C}{4} absorption lines at
  \dtra\ $\sim$ 500~kpc are about one order of magnitude smaller than
  those at \dtra\ $\sim$ 100~kpc \citep[e.g.,][]{tur14,tur_phd}.  The
  CGM absorption strength also depends on physical properties of host
  galaxies such as luminosity, star formation rate, stellar mass, and
  a number density of galaxy in Mpc scale around that
  \citep{che01,bor14,bur16}.}
\item{Our sample is heterogeneous in terms of spectral resolution
  (i.e., $\lambda / \Delta \lambda$ $\sim$ 1000 -- 27000) and data quality 
  (i.e., a wide range of S/N ratio).  It also should be noted that strong absorption
  lines (\rew\ $\sim$ 1 -- 2 \AA) in our sample have two possible origins:
  a ``single" dense gas system and a clustering of gas clumps of narrow 
  absorption lines, although we cannot separate them into the two groups 
  with our low-resolution spectra.  It is highly required to perform the same 
  analyses based on homogeneous (and higher resolution) spectra taken 
  with same telescope and instrument with a specific observing configuration.}
\end{itemize}

\section{SUMMARY}
We collected spectra of 13 gravitationally lensed quasars from SDSS
Quasar Lens Search catalog as well as from the literature, and
investigated the fractional equivalent width difference \dew\ and 1on
ratio \rone\ as a function of the physical separation in the
transverse direction \dtra. We also constructed simple models with
five parameters to reproduce the observed results based on 293 metal
absorption lines to investigate the internal structure of the CGM. Our
main results are as follows.

\begin{itemize}
\item Correlation coefficients between absorption strength (i.e.,
  \rew) along sightline pairs are almost same for high-ionized lines
  (e.g., \ion{C}{4}) and low-ionized lines (e.g., \ion{Mg}{2}),
  although the latter tends to have large scatter at \rew\ $<$ 1 \AA.
\item Typical size of high-ionized absorbers is probably larger than
  that of low-ionized absorbers because the former has small 1on ratio
  (\rone\ $\sim$ 2\%) only at larger physical distance between
  sightlines of lensed images (\dtra\ $\sim$ 10~kpc) while the latter
  has \rone\ $\sim$ 16\% at the smaller scale of \dtra\ $\sim$ 1~kpc.
\item Both high and low-ionized absorbers have almost same values of
  the fractional equivalent width difference \dew\ $\sim$ 0.2 for a
  wide range of sightline separations \dtra\ $\sim$ 0.1 -- 10~kpc,
  although the latter has a larger scatter of \dew.
\item We constructed simple models for the CGM using five parameters;
  equivalent width distribution as a function of the radius from a
  spherical cloud center ($\ew(r)$), a size (diameter) of each
  spherical cloud ($d$), an overall size of the CGM ($L$), an
  intensity of equivalent width by diffuse gas ($\ew_{\rm diff}$), and
  a covering factor (\cf).
\item Acceptable ranges of these parameters for high-ions are $d$
  $\leq$ 1~kpc, $L$ $>$ 100 -- 500~kpc ($L$ $>$ 500~kpc is more
  reasonable), \cf\ = 2, with no diffuse gas or $d$ $\leq$ 0.5~kpc,
  $L$ $>$ 100 -- 500~kpc, \cf\ = 1, with a diffuse gas whose
  equivalent width is about a half the peak value of each spherical
  cloud. Those for low-ions are $d$ $\leq$ 1~kpc, $L$ $\geq$
  500~kpc, \cf\ = 1.5, with no diffuse gas. Both the elliptical and
  the inverse-proportional functions are acceptable, although the
  former gives a better result.
\end{itemize}

Comparing the models and the observations, we placed constraints on
the internal structure of the CGM with acceptable ranges for five
parameters.  Our best model gives a picture of the CGM that is similar
to those in the literature: low-ionized absorbers have a complex
internal structure consisting of a large number of small-scale clouds,
and they are embedded in the diffusely distributed high-ionized
regions.

Our results suggest that more detailed analysis using larger samples
taken with same telescope/instrument is important to strengthen our
conclusion.  More lensed quasars will be discovered by deep imaging
surveys like the Hyper Suprime-Cam Subaru Strategic Program (HSC-SSP;
\citet{aih17}) with follow-up spectroscopic observations. If
absorption lines are resolved completely in high-resolution spectra,
we can further discuss the internal structure of the CGM based on
column density (rather than equivalent width) and velocity structure
by applying Voigt profile fitting. We also need to perform deep
imaging observations to detect host galaxies of our sample absorption
systems to investigate the physical properties of the CGM as a
function of the impact parameter from host galaxies of absorbers. By
doing so, we will be able to discuss a volume number density of gas
clouds as a function of the distance from the host galaxies to improve
our current models.

\acknowledgments
We would like to thank the anonymous referee for
valuable comments that helped us to improve the paper.  We also would
like to thank Christopher Churchill for providing us with the software
package, {\sc search}.  We also would like to Masami Ouchi, Akio
Inoue, and Shiro Mukae for their helpful comments and discussions.
The research was supported by the Japan Society for the Promotion of
Science through Grant-in-Aid for Scientific Research 15K05020 and
partially supported by MEXT Grant-in-Aid for Scientific Research on
Innovative Areas (No.~15H05894). NK acknowledges supports from the
JSPS grant 15H03645.  KO acknowledges supports from the JSPS grant
16K05299.

\end{document}